\documentclass[10pt,twocolumn,preprintnumbers,amsmath,amssymb,nofootinbib,superscriptaddress,aps,prd,numberedappendix]{openjournal}
\usepackage{newtxtext,newtxmath}
\usepackage{amsmath}
\usepackage{amssymb}
\usepackage{graphicx}
\usepackage{dcolumn}
\usepackage{bm}
\usepackage[dvipsnames]{xcolor}
\usepackage{xspace}
\usepackage{ulem}
\usepackage{lipsum}
\usepackage{xfrac}
\usepackage[colorlinks=true, allcolors=blue]{hyperref}

\usepackage{booktabs}
\usepackage{braket}
\usepackage{tabularx}

\begin{document}

\title{Recovering 21cm monopole signals without smoothness}

\author{Rugved Pund$^{1,2*}$}
\author{An\v ze Slosar$^{1,2}$}
\author{Aaron Parsons$^3$}
\email{$^*$rugved.pund@stonybrook.edu}
\affiliation{$^1$Physics and Astronomy Department, Stony Brook University, Stony Brook, NY 11794}
\affiliation{$^2$Physics Department, Brookhaven National Laboratory, Upton, NY 11973, USA}
\affiliation{$^3$Department of Astronomy, University of California, Berkeley, Berkeley, CA 94720, USA}
\date{\today}

\begin{abstract}
We expect the monopole signal at the lowest frequencies below $100\,$MHz to be composed of two components: the deep Rayleigh-Jeans tail of the cosmic microwave background and two distinct features: the dark ages trough at $\sim 17\,$MHz and the cosmic dawn trough at $\sim 75\,$Mhz. These are hidden under orders of magnitude brighter foregrounds whose emission is approximately a power-law with a spectral index $\approx -2.5$. It is usually assumed that monopole signals of interest are separable from foregrounds based on spectral smoothness. We argue that this is a difficult approach and likely impossible for the Dark Ages trough. Instead, we suggest that the fluctuations in the foreground emission around the sky should be used to build a model distribution of possible shapes of foregrounds, which can be used to constrain the presence of a monopole signal. We implement this idea using normalizing flows and show that this technique allows for efficient unsupervised detection of the amplitude, width, and center of the Dark Ages trough as well as the Rayleigh-Jeans tail of the cosmic microwave background for a sufficiently sensitive experiment. We show that achromatic and smooth response significantly helps with foreground separation. We discuss the limitations of the inherent assumptions in this method and the impact on the design of future low-frequency experiments.
\end{abstract}
\maketitle

\section{Introduction}

\begin{figure}
    \centering
    \includegraphics[width=\linewidth]{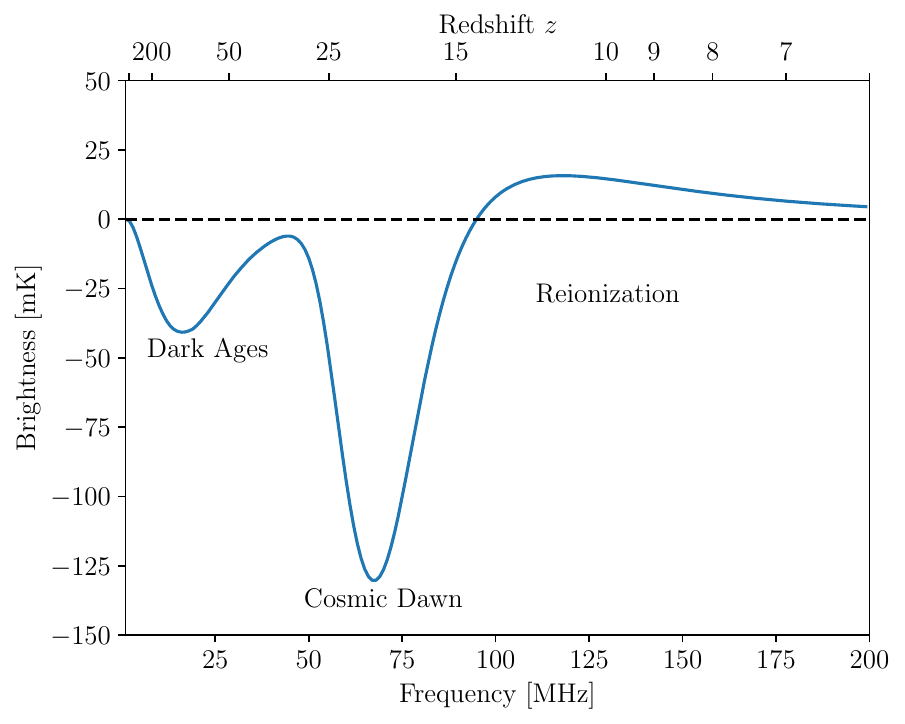}
    \caption{Evolution of the 21cm cosmic hydrogen signal with redshift and frequency}
    \label{fig:21cm}
\end{figure}

After the epoch of recombination at redshift around $z\sim 1150$ and before the universe is completely reionized at redshifts $z\lesssim 6$, the intergalactic medium (IGM) is filled with neutral hydrogen. During this period, the hyperfine spin-flip transitions in the hydrogen gas absorb and emit the characteristic $21\,$cm radiation at $1420\,$MHz, which is predicted to be observable against cosmic microwave background emission at radio frequencies below $\nu \lesssim 200\,$MHz due to cosmological redshift since. This period in the evolution of the universe can be approximately divided into three chapters: the Dark Ages, before the advent of the first stars, where the neutral hydrogen is largely in its pristine primordial state, cooling adiabatically with the expansion of the universe and weak Compton scattering with the CMB; the Cosmic Dawn, when the first sources of light other than the CMB begin forming, and start to heat and photo-ionize the neutral primordial medium, eventually culminating in reionization; and the Post-Reionization evolution when hydrogen recombines in galaxies to form a biased tracer of the cosmic filaments of large scale structure. Each episode is associated with a mean absorption profile as illustrated in Figure \ref{fig:21cm}. For an in-depth discussion of these effects, see \cite{furlanettoCosmologyLowFrequencies2006,pritchard_21-cm_2012,furlanetto_fundamentals_2019}.

Measurements of the 21\,cm line at high redshift would be transformative for our understanding of the evolution of our universe. Many current experiments seek to measure the global 21\,cm signal during the Cosmic Dawn (e.g., 
CTP \citep{Nhan:2017}, EDGES \citep{Bowman:2018}, MIST \citep{Monsalve:2020}, SARAS \citep{Singh:2017}, REACH \citep{deLeraAcedo:2019}, BIGHORNS \citep{Solokowski:2015}, LEDA \citep{Price:2018}, PRIZM \citep{Philip:2019}, DAPPER \citep{Burns:2019}, and SCIHI \citep{Voytek:2014}),
%Experiment to Detect the Global Epoch-of-Reionization Signature (EDGES; Bowman et al. 2008; Bowman & Rogers 2010; Monsalve et al. 2017), the Sonda Cosmológica de las Islas para la Detecciónde Hidrógeno Neutro (SCI-HI; Voytek et al. 2014), the Probing Radio Intensity at high-z from Marion (PRIzM; Philip et al. 2019), the Shaped Antenna measurement of the background RAdio Spectrum (SARAS; Patra et al. 2013; Singh et al. 2017, 2018), the Cosmic Twilight Polarimeter (CTP; Nhan et al. 2019), the Broadband Instrument for Global Hydrogen Reionization Signal (BIGHORNS; Sokolowski et al. 2015), the Large-Aperture Experiment to Detect the Dark Age (LEDA; Bernardi et al. 2016; Bernardi 2018; Price et al. 2018), and the Radio Experiment for the Analysis of Cosmic Hydrogen (REACH; de Lera Acedo 2019). 
as well as the power spectrum of fluctuations about the mean during the Cosmic Dawn (e.g. 
MWA \citep{Bowman:2013, Beardsley:2016, Barry:2019}, LOFAR \citep{Patil:2017}, PAPER \citep{Parsons:2014}, LWA \citep{Eastwood:2019}, HERA \citep{DeBoer:2017}) and the later epoch of cosmic expansion (e.g., CHIME \citep{Amiri_2023}, HIRAX \citep{Newburg_2016}).

The mean signal and the fluctuations provide unique and complementary information for understanding the evolution of cosmic structure \citep{Liu:2016}. While next-generation experiments on the ground aim to expand sensitivity and improve control of systematics, renewed attention is being paid to the possibility of measuring the Dark Ages signal from outside of the Earth's ionosphere 
\citep{baleLuSEENightLunar2023,raoPRATUSHExperimentConcept2023,jiaScientificObjectivesPayloads2018,Boonstra:2010,Chen:2019,Burns:2019,Shi:2021,Zhu_2021,Shi_2022,kleinwoltAstronomicalLunarObservatory2024}
%Discovering the Sky at the Longest wavelength (DSL; also known as Hongmeng in Chinese, Chen et al. 2019, 2020), the Dark Ages Polarimetry PathfindER (DAPPER; Tauscher et al. 2018a) and its precursor the Dark Ages Radio Explorer (DARE, Burns et al. 2017), and the Farside Array for Radio Science Investigations of the Dark ages and Exoplanets (FARSIDE; Burns et al. 2019, 2021).

The \emph{Dark Ages} absorption trough is of depth $\sim40\,$mK and is centered at around $\nu \sim 17\,$MHz and has a width of $\Delta \nu \sim 8\,$MHz \citep{furlanettoCosmologyLowFrequencies2006}. Its precise shape can be calculated from the first principles assuming only the standard cosmology (i.e., linearized general relativity, thermodynamics, and atomic physics, but not astrophysics) and is very sensitive to any injection or sink of energy in the primordial medium. It has never been measured and will very likely require a space-based observatory.

The \emph{Cosmic Dawn} absorption trough is of depth $\sim140\,$mK and is centered at around $\nu \sim 75\,$MHz and has a width of $\Delta \nu \sim 10\,$MHz \citep{furlanettoCosmologyLowFrequencies2006}. Its shape is sensitive to the physics of first stars and the process of reionization. Several active experimental programs are attempting to constrain this signal, with the EDGES team \citep{bowman_absorption_2018} reporting evidence at $78\pm1\,$MHz for a symmetric U-shaped profile of amplitude $500^{+500}_{-200}\,$mK. However, this possible detection could arise from modelling or as instrumental systematics \citep{1805.01421, bradley_ground_2019,simsTestingCalibrationSystematics2020}, and the SARAS3 team has reported measurements that are inconsistent with the EDGES model, rejecting them with 95.3\% confidence \citep{singh_detection_2021}.

Finally, the same frequencies of interest should also contain the deep Rayleigh-Jeans tail of the cosmic microwave background. Detection of this background would be of interest from both theoretical perspective \citep{1803.07048,2206.07713,2209.09063} as well as clarification on the recent excess monopole radiation at the very low frequencies \citep{1804.08581,2011ApJ...734....5F}.

The main difficulty in experimental detection of the signal is the existence of bright foregrounds. These foregrounds are approximately smooth power-laws in frequency with a spectral index of $\sim2.5$ that varies to a small extent depending on spatial directions. Due to a lack of a full a-priori model of the sky and being orders of magnitude brighter than the signal of interest, $21\, $cm signal separation from the foregrounds continues to be a challenge. Instrument systematics like chromaticity, calibration errors, and beam shape can further complicate the separation and there have been suggestions of several Bayesian analysis pipelines to address them, for example, see \cite{ansteyGeneralBayesianFramework2021, ansteyInformingAntennaDesign2021, acedoREACHRadiometerDetecting2022, paganoGeneralBayesianFramework2023, ansteyUseTimeDependent2023, saxenaSkyaveraged21cmSignal2023}. The foreground model itself in several of these approaches has been broadly based on the assumption of smoothness, for example, maximally smooth functions \citep{raoDetectionSpectralRipples2015, sathyanarayanaraoModelingRadioForeground2017, bevinsMaxsmoothRapidMaximally2021a} and physically motivated models that use polynomials or log-polynomials have been used in the analysis of EDGES and SARAS3 data \citep{bowman_absorption_2018, singh_detection_2021}.

In Section \ref{sec:separation}, we use a toy model to argue that residuals from smooth foreground models could potentially be misleading. Moreover, the assumption of smoothness would likely make separating the Dark Ages signal nearly impossible, given its smaller depth and larger width than the Cosmic Dawn signal. In Section \ref{sec:method}, we propose a different method for separating the signal of interest from the foregrounds with its own unique assumptions. In Sections \ref{sec:implementation} and \ref{sec:results}, we discuss its implementation and results when applied to synthetic datasets. We conclude in Section \ref{sec:conclusions}.

\section{Separating foregrounds}
\label{sec:separation}

\newcommand{\vxh}{\hat{\mathbf{x}}}
\newcommand{\Tfg}{\bar{T}_{\rm fg}}
\newcommand{\DTfg}{\Delta T_{\rm fg}}
\newcommand{\Ts}{T_{\rm s}}
\newcommand{\Pfg}{P_{\rm fg}}

Separating foregrounds from the signal is the foundational problem of 21\,cm cosmology. We can write the sky temperature in direction $\vxh$ and frequency $\nu$ as
\begin{equation}
     T(\vxh,\nu) = \Ts(\nu) + T_{\rm fg}(\vxh,\nu) = \Ts(\nu) + \Tfg(\nu) + \DTfg(\vxh,\nu),
\end{equation}
where we have split the signal into monopole signal of interest $\Ts(\nu)$ and the foreground temperature $T_{\rm fg}(\vxh,\nu)$.

In this treatment, we ignore spatial fluctuations in the cosmological signal and focus solely on the monopole signal. The fluctuations in the foreground are typically several orders of magnitude brighter than the signal. We further split them into the mean foreground monopole $\Tfg(\nu)$ and variations around this mean $\DTfg(\vxh,\nu)$. By definition, then, these fluctuations must average to zero $\left<\DTfg(\vxh,\nu) \right>_{\vxh}=0$, where the average is defined as usual over a sufficiently large observational patch. 

From now on, we will drop the explicit frequency dependence in our notation and assume that our measurements have an implicit frequency dependence of 50 bins of 1MHz width each. We will still explicitly include spatial dependence to distinguish between the monopole signals (which are functions of just frequency) and quantities that vary both with frequency and position on the sky.

It is evident from the above that $\Ts$ and $\Tfg$ are perfectly degenerate in any observation unless we make further assumptions that enable separation. 

\subsection{Smoothness separation}

We can attempt to separate the foreground emission from the signal by assuming that the mean foregrounds are much smoother than the signal in the frequency direction. The mean foregrounds vary primarily as a smooth power law, while the cosmological signals have one or several closely spaced inflection points. Schematically, a high-pass filter applied to the $\Ts+\Tfg$ will filter out the $\Ts$ and result in $\approx \Tfg$. This approach is based on intuition from spectroscopy, where the equivalent line widths can be measured quite accurately, even for spectra where broadband spectro-photometry is relatively poor. 

Quantitatively, suppose we have a narrow feature on top of a slowly varying background where the cosmological signal is larger than the typical variation of the background over the width of the feature. In that case, one should be able to extract it cleanly. For power-law backgrounds, $B(\nu)\sim A \nu^{-\alpha}$ with an absorption feature of amplitude $S$ imprinted over some frequency range $\Delta \nu$, typical variation of the background over the width of the feature can be estimated as
\begin{equation}
    \frac{\Delta B}{B} \sim \frac{\Delta \nu}{\nu} A
\end{equation}
(i.e., for locally power-law shaped backgrounds, it will be of this form with the spectral index as a prefactor.). This also determines the typical uncertainty in the background determination over the width of the feature. To cleanly separate the feature from the smooth foreground, we require that this uncertainty be subdominant to the feature itself, i.e., the signal-to-background ratio $S/A$ should be larger than the fractional background uncertainty
\begin{equation}
    \frac{S}{A} > \frac{\Delta \nu}{\nu} A.
\end{equation}
For the Cosmic Dawn signal, $S/A\sim 10^{-3}$ and for the Dark Ages signal $S/A\sim 10^{-5}$. On the other hand, we have $\Delta \nu / \nu \sim 0.1$ and $\sim 1$, respectively. 
From this, we see that this smoothness separation condition has already been violated for the Cosmic Dawn feature. Therefore, it is not surprising that this detection's statistical significance has been challenged in the literature \citep{1805.01421,bradley_ground_2019}. In general, the violation of this requirement makes the significance of a detection sensitive to the exact modelling of the smooth foregrounds. The condition is even more badly violated for the Dark Ages feature.

\begin{figure}
     \includegraphics[width=\linewidth]{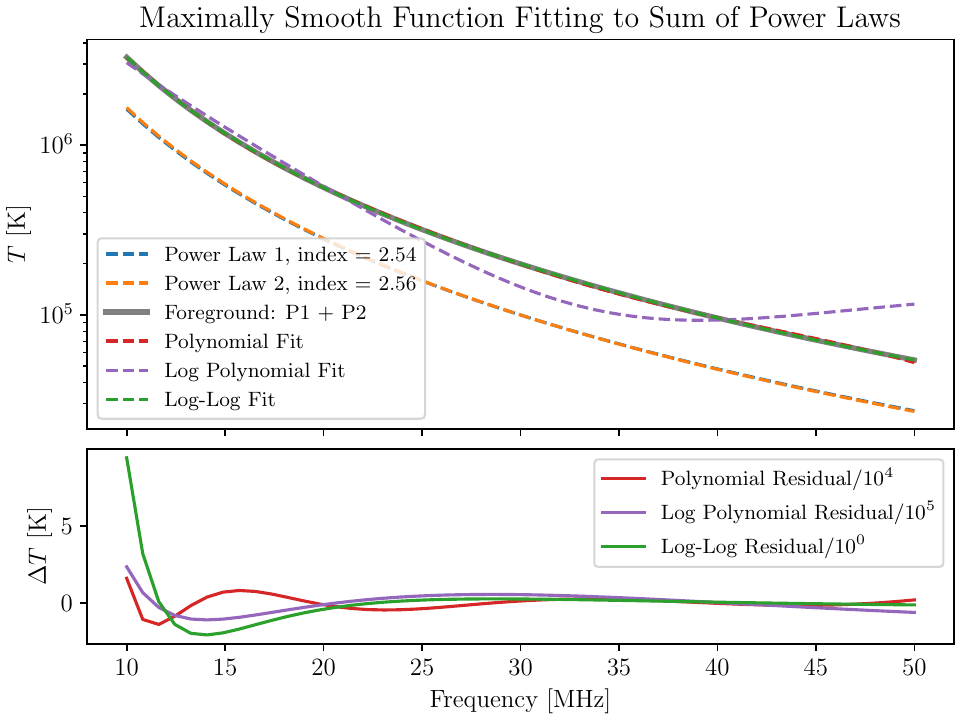}
     \caption{Toy Example: A sum of two power laws with spectral indices of -2.54 (blue) and -2.56 (orange) is fit with maximally smooth functions of 15 polynomials (red), log-polynomials (purple) and log-log polynomials (green). Residuals are shown in the bottom panel.
     \label{fig:toy}}
\end{figure}

To further illustrate this point on a concrete example, we consider the following toy model: suppose there are two patches of the sky, each emitting a pure power-law emission with spectral indices -2.54 and -2.56, respectively, and contributing equally to the total emission. We fit the total mean foreground with a maximally smooth function (MSF) using the \texttt{maxsmooth} package (see \cite{bevinsMaxsmoothRapidMaximally2021a}) with the basis chosen as polynomials, log-polynomials, and loglog-polynomials as the basis of maximum order 15 in the expansion for the fit. As shown in Figure \ref{fig:toy}, the best-fits still show residuals of various amplitudes that can be confused as a false detection. (We note that the log-polynomials with order \texttt{N=15} have a markedly poor fit despite the large order and need significantly more terms that take an unreasonably long time to converge). The crucial point is that residuals have structures that vary over similar scales as the actual signal, while the amplitude is still over an order of magnitude larger than the signal of interest. Both the \texttt{maxsmooth} paper and the earlier implementations in \cite{raoDetectionSpectralRipples2015,sathyanarayanaraoModelingRadioForeground2017} recommend several diagnostics for improving and validating the fitting routines that could be implemented in this case. In particular, the use of Bayesian Information Criterion and the change in Bayesian evidence after jointly fitting the signal and foregrounds, can be used to evaluate the goodness of fit and are recommended in the literature to evaluate relative modeling systematics. Still, the need for additional diagnostics demonstrates the central issue of separation based on foreground smoothness.

In reality, where there are multiple power-law like contributions to the foregrounds, along with instrumental noise and systematics, the residuals after fitting a suitable number of appropriately chosen smooth functions could incorrectly indicate the presence of a strong 21cm signal (if the fitting model is insufficiently flexible) or a nearly zero 21cm signal (if the fitting model is overly flexible). Fundamentally, unless one has a physical model of the sky that is accurate at the level below the amplitude of target features, there is no robust way of distinguishing the under-fit and over-fit scenarios.

The situation is somewhat less drastic in the case of Cosmic Dawn since the foregrounds are both less bright and the peak sharper. At the same time, the sky might also be more complicated than we think it is, so smoothness-based separation techniques remain a challenge.

\subsection{Explicit foreground modelling}

An alternative to smoothness-based separation is to model spectral degrees of freedom in the foreground. 
Such a model can have a large number of free parameters that can be determined from the data, but in this approach, the functional form of the model is determined in advance or derived from a large set of realistic simulations called training datasets in this context. See \cite{1711.03173, 2011.01242} for examples of this approach in action. 
However, to reach the precision required to subtract the foregrounds at $10^{-4}$ to $10^{-6}$ level required by the data is a daunting task that will likely need a very large number of parameters, perhaps exceeding a number that can be efficiently sampled using traditional Monte Carlo Markov Chain approaches.

\subsection{Data-driven foreground modelling}

An alternative possibility is to infer the model for the foregrounds from the data itself. Any variations on the sky from one point to another, at least in the limit of an ideal instrument, are caused by the foreground variations and not the monopole component. Given a model of these variations and their relation with the foreground-only monopole, we can then use the model to constrain the presence of the non-foreground monopole.

The approach we take in this paper is to replace the assumption of smoothness with an assumption based on the observation that the foreground fluctuations from the mean, $\DTfg(\vxh,\nu)$, across the sky at any particular frequency are of the same order of magnitude if not larger than the mean foreground $\Tfg(\nu)$. In terms of the probability distribution of the foreground fluctuations, the mean foreground  $\Tfg$ at a particular frequency belongs to the distribution of foreground fluctuations $\DTfg(\vxh)$ across the sky at the same frequency. If $\mathbb{F}$ is the set of all possible deviations from the mean foreground, i.e., $\mathbb{F}(\nu) = \{ \DTfg(\vxh,\nu) | \vxh \in {\rm sky} \}$, then $\Tfg \in \mathbb{F}$. This is admittedly a strong assumption but is rooted in the physical intuition that foreground-generating processes are local. As an example, consider a sky that is generated as a combination of two power laws, as in the full sky version of our toy example earlier (Figure \ref{fig:toy})
\begin{align}
    T_{\rm toy}(\vxh) = A(\vxh) \left(\frac{\nu}{\nu_A}\right)^{-2.54} &+ B(\vxh) \left(\frac{\nu}{\nu_B}\right)^{-2.56} \\
    \bar{T}_{\rm toy} = \bar{A} \left(\frac{\nu}{\nu_A}\right)^{-2.54} &+ \bar{B} \left(\frac{\nu}{\nu_B}\right)^{-2.56} \\
    \Delta T_{\rm toy}(\vxh) = (A(\vxh) - \bar{A}) \left(\frac{\nu}{\nu_A}\right)^{-2.54} &+ (B(\vxh) - \bar{B}) \left(\frac{\nu}{\nu_B}\right)^{-2.56} \\
\end{align}
In this case, the set $\mathbb{F}$ is composed of linear combinations of power laws with exponents $-2.54$ and $-2.56$ and indeed $\Tfg$ is in $\mathbb{F}$, while some general function of temperature is not. 

In summary, we propose that the fluctuations in the foreground emission across the sky should be used to build a model distribution of possible shapes of the mean foregrounds, which can be used to constrain the presence of a 21cm monopole signal. This approach is agnostic to the exact shape of the monopole signal since the model distribution is built entirely from the fluctuations around the mean sky, which subtracts the monopole signal by construction from the data.

In the following sections, we will develop a probabilistic framework for constraining the signal of interest $\Ts$ based on a generative model for $\mathbb{F}$ of the foregrounds.

\section{Method}
\label{sec:method}
\newcommand{\T}{\mathbf{T}_{\rm fg}}
\newcommand{\DT}{\Delta\T}
\newcommand{\F}{\mathbf{F}_{\rm fg}}

For a start, we assume a perfect instrument without any calibration errors or gain fluctuations. In this case, assuming temperature data from the beam-convolved sky $T(\vxh)$, we can calculate the mean sky monopole data,
\begin{equation}
    T_m = \left<T(\vxh)\right> = \Tfg+\Ts
\end{equation}
and fluctuations from the monopole,
\begin{equation}
    \DTfg(\vxh) = T(\vxh) - T_m,
\end{equation}
noting that the mean sky monopole contains both the mean foreground $\Tfg$ and the signal of interest $\Ts$, and the foreground fluctuations contain no contribution from the signal of interest.

A careful reader might protest that data from the sky $T(\vxh)$ depends on the shape of the telescope's beam. This is of course true, but crucially, while the statistical model of the sky fluctuations will depend on the shape of the beam, the sky-averaged quantities $T_m$ remain unaffected. 

Starting with the distribution of the foreground fluctuations based on the samples $\mathbb{F} = \{\DTfg(\vxh)\}$ across the sky, we build a probabilistic model $\Pfg$ of the distribution $\mathbb{F}$. This allows us to calculate explicitly the likelihood function for the distribution $\Pfg(T)$ for any given $T(\nu)$.

Ultimately, we are interested in the probability of a signal being present in the monopole $P(\Ts | T_m)$. This can be written in terms of the likelihood of the foreground model simply as 
\begin{equation}
    P(\Ts | T_m) = \Pfg(T_m - \Ts)
    \label{eq:base}
\end{equation}
since the foreground model is constructed from the fluctuations around the mean sky monopole. Thus, the model likelihood of the signal $\Pfg(T_m - \Ts)$ will be maximized if the correct template $\Ts$ for the signal is subtracted from the sky monopole,
\begin{equation}
    P(\Ts|T_m) = \Pfg(T_m - \Ts) \xrightarrow[\Ts \rightarrow \Ts^{\rm true}]{} \Pfg (T_m-\Ts^{\rm true})
\end{equation}
In other words, we can detect the signal of interest $\Ts^{\rm true}$ in the sky monopole by maximizing the likelihood of the foreground in the sky monopole after subtracting signal templates with the right amplitude, width, etc.

This also allows us to find a maximum likelihood template for the signal given a foreground model as
\begin{equation}
    \Ts^{\rm best} = \max_{\{\Ts'\}} \Big[ \Pfg\left(T_m-\Ts'\right) \Big] % \xrightarrow[\text{correct } \Ts]{} 
\end{equation}
However, we shall see that some components of the signal of interest are significantly dominated by the foregrounds, making data-driven methods unable to recover the full unbiased signal. Thus, it only makes sense to compare the relative likelihoods of one template signal to another.

Our implementation proceeds as follows. We first calculate the PCA basis of $\DTfg$ and then rotate the full set of $\DTfg$ into it. We then apply a normalizing flow transformation to map these vectors into a Gaussian distribution. The Jacobian of the inverse map then provides us a probabilistic model for $\DTfg$, which we can use in a probabilistic model for $\Ts$ in Equation (\ref{eq:base}).

\section{Synthetic Data and Implementation Details}
\label{sec:implementation}

\subsection{Foreground sky maps and model observations}

To generate our synthetic data, for the Dark Ages signal, we use the Ultra Low-wavelength Sky with Absorption (ULSA) model from \cite{cong_ultra-long_2021}, and for the Cosmic Dawn signal, the improved Global Sky Model (GSM16) model from \cite{zhengImprovedModelDiffuse2017}. 

The ULSA model is the current state-of-the-art model for the low-frequency sky. While anchored in high-frequency data, most importantly the Haslam 402\,MHz map, it is valid down to $\sim1$\,MHz, features direction and frequency-dependent spectral indices, accounts for free-free absorption by thermal electrons in the galaxy, etc. The free-free absorption from the galaxy's warm ionized medium causes a notable turnover effect in the $\lesssim 10$MHz frequency range, a significant departure from the power-law radio background model.  

The GSM16 model is the improved Global Sky Model of the diffuse galactic radio emission from 10\,MHz to 5\,THz from \cite{zhengImprovedModelDiffuse2017}, based on the original GSM2008 from \cite{deoliveira-costaModelDiffuseGalactic2008}. The GSM16 model includes the enhanced Haslam map, and 29 other sky maps from various radio experiments, and implements an improved PCA algorithm that extends the original GSM algorithm based on interpolating the best-fit PCA components in frequency space between 50-100\,MHz frequencies relevant for the Cosmic Dawn signal.

While it is known that these foreground maps are not an entirely accurate description of reality, since they are consistent with measurements, they currently offer the highest level of realism. They are sufficiently complex to be unlikely to give misleadingly optimistic results.

We generate maps with the \texttt{HEALPix} \citep{gorski_healpix_2005} resolution \texttt{NSIDE}=128 for both foreground models using the codes provided with their respective papers.  We sample frequencies in 1MHz bins in the range 1-50\,MHz for the ULSA model and 50-100\,MHz for the GSM16 model. 

To observe these maps, we assume Gaussian beams with $\sigma=2^\circ$, $4^\circ$, and $6^\circ$ at $10\,$MHz.  In the frequency direction, we either keep the beam size constant (achromatic) or scale it with the inverse of the observation frequency (chromatic). We will refer to beams by their size at $10\,$MHz, and chromaticity. These are again not very realistic, but sufficient for our exploratory investigations regarding the limits of our foreground modelling approach, effects of chromaticity, calibration systematics and ways to mitigate them using multi-scale combined maps. In future studies, we hope to address the impact of realistic beams typical for sky-averaged measurements.

\subsection{Noise model}

We proceed by adding measurement noise. For our analysis, we consider a noise profile for RMS noise, as given by the radiometer equation, that is proportional to the temperature of the sky in each frequency bin for a given SNR. We assume that the noise is uncorrelated between frequency bins and between pixels. The noise level is given by the radiometer equation,
\begin{equation}
\sigma_{\rm noise} = \frac{\Tfg}{\rm SNR},
\end{equation}
where 
\begin{equation}
    {\rm SNR} = \sqrt{N_d B t}
\end{equation}
corresponds to the map-level signal-to-noise ratio for $N_d$ detectors operating over bandwith $B$ for time $t$ and $\Tfg$ is the mean foreground temperature in each frequency bin. We label the noise levels by their map-level SNR. Of course, the level of noise in each direction is then larger by the square root of the number of pixels. 

Compared with the reach of existing and planned experiments for Cosmic Dawn, where the brightest foregrounds are of the order $10^4$\,K, an SNR of $10^5$ corresponds to a map-level RMS noise of 100\,mK. Similarly, the SNR  $10^6$ and $10^7$ correspond to map-level RMS noise of 10\,mK and 1\,mK, respectively. We note that this also agrees with the expectations of the REACH team's 250\,mK, 25\,mK, and 5\,mK noise levels for the pessimistic, realistic, and optimistic scenarios \citep{acedoREACHRadiometerDetecting2022}.

For the Dark Ages signal with the brightest foregrounds of the order $10^7$\,K, an SNR of $10^8$ corresponds to a map-level RMS noise of 100\,mK. Similarly, the SNR  $10^9$ and $10^{10}$ correspond to map-level RMS noise of 10\,mK and 1\,mK respectively. While practically achieving these noise levels is a challenge, they are useful for understanding the contexts of our foreground modelling approach.

\begin{table}
\centering
\caption{SNR Requirements for Observational Scenarios}
\label{tab:snr_scenarios}
\begin{tabularx}{0.8\columnwidth}{ccc}
\toprule
\textbf{Scenario} & \textbf{SNR Dark Ages} & \textbf{SNR Cosmic Dawn}  \\
\midrule                                                              
Optimistic        & $10^{10}$              & $10^7$                    \\ 
Realistic         & $10^{9}$               & $10^6$                    \\
Pessimistic       & $10^8$                 & $10^5$                    \\
\bottomrule
\end{tabularx}
\end{table}

\subsection{Galaxy cuts}

Finally, in all our investigations, we remove the very bright regions around the galactic plane by imposing the galactic cut of $\pm 20^\circ$ in galactic latitude. We have observed that including the galaxy in the data severely limits our ability to detect the signal (see Appendix \ref{app:galcut} for a full comparison). However, the precise size of the galactic cut is not crucial. 

In the real world, observing campaigns that target times when the hottest parts of the galactic plane are below the horizon are expected to be more successful in detecting the signal of interest \citep{parsonsPrecisionArrayProbing2010, deboerHydrogenEpochReionization2017}. While the galactic cut at the map level does not reflect the reality of typical instruments that have to deal with the galaxy in their beam, we have found that it is a useful simplification for this study. A study involving realistic beams for the instruments and the differences in constraints between galaxy-up vs. galaxy-down observations during the instrument drift scans is being planned for investigation as a part of upcoming work. 

\subsection{Rotation into PCA basis}

\begin{figure*}
    \includegraphics[width=0.49\linewidth]{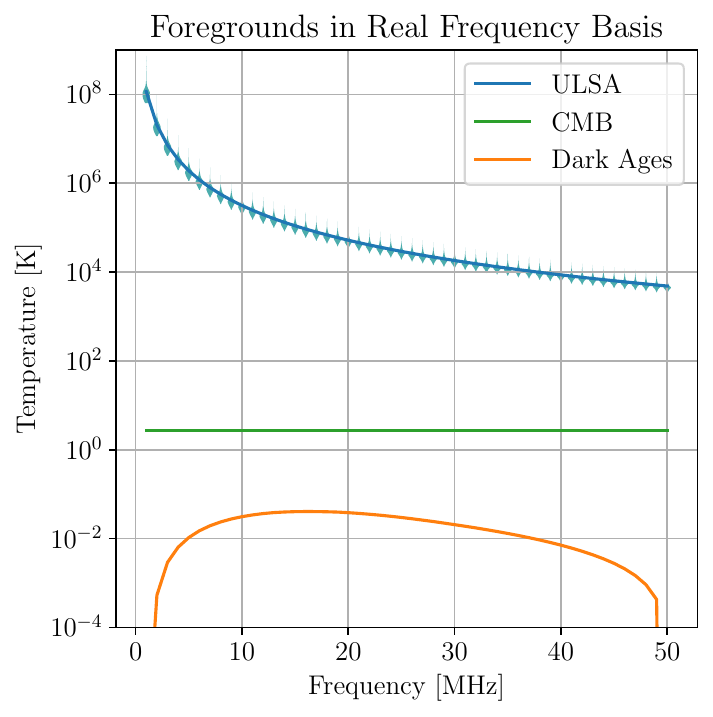}
    \includegraphics[width=0.49\linewidth]{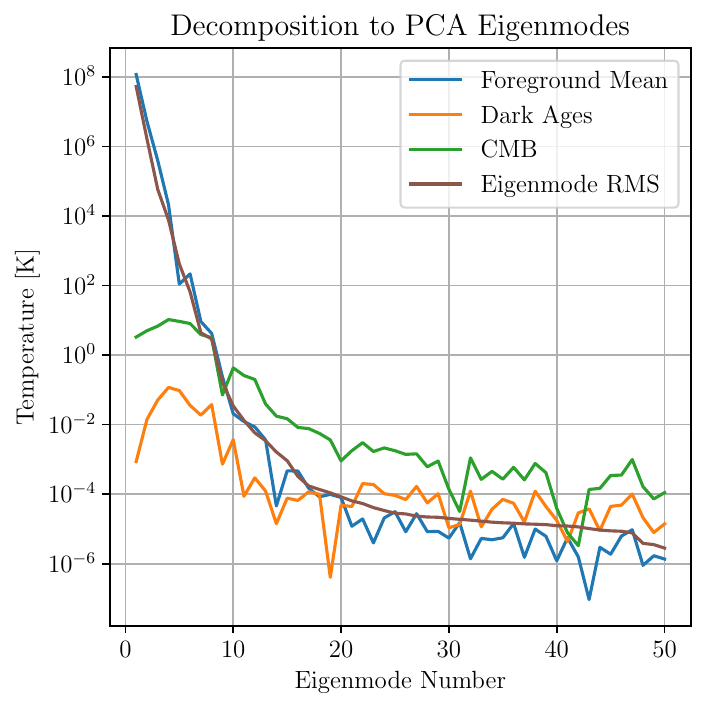}
    \includegraphics[width=1.0\linewidth]{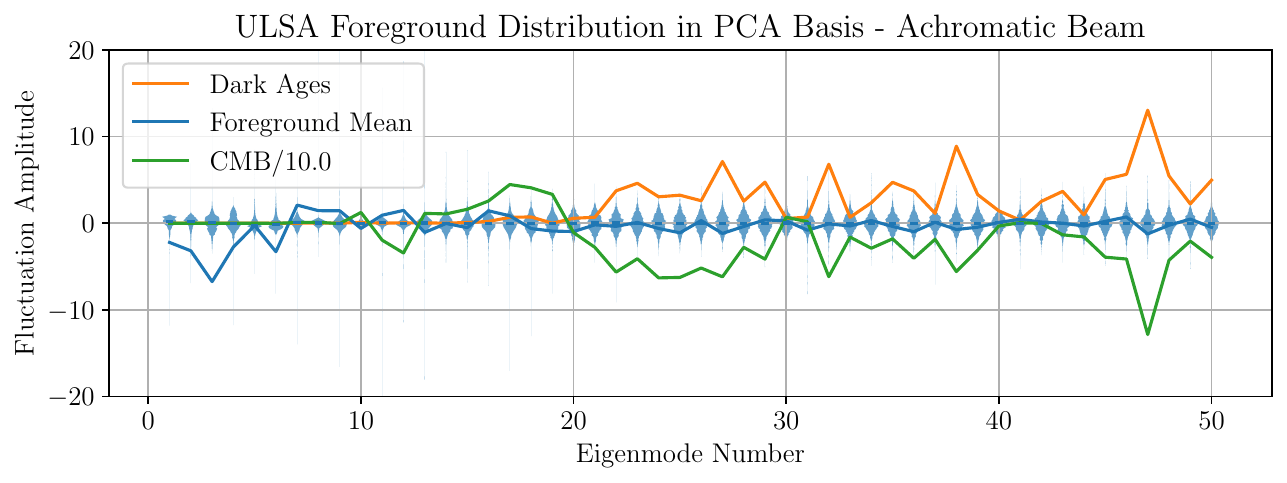}
\caption{This figure shows the separation of foregrounds and the Dark Ages signal in different spaces for the ULSA foreground model. \emph{Top Left}: Separation in the original frequency space using an achromatic $2^\circ$ beam. The blue distribution violins and the solid line describe the foreground signal distribution and mean (on the cut sky), the green line denotes the CMB signal, and the orange solid line denotes the Dark Ages signal. Note the extremely large dynamic range on the vertical axis. \emph{Top Right}: The amplitude of the same signals in the foreground PCA basis. The first ten principal components absorb about 8 orders of magnitude of the signal. The brown solid line denotes the RMS of the corresponding eigenmode. \emph{Bottom Panel} Signal amplitudes scaled by the RMS of the corresponding eigenmode. The basis vectors have been multiplied by $-1$ to make all the Dark Ages contributions positive.\label{fig:pca}}
\end{figure*}

\begin{figure*}
    \includegraphics[width=0.49\linewidth]{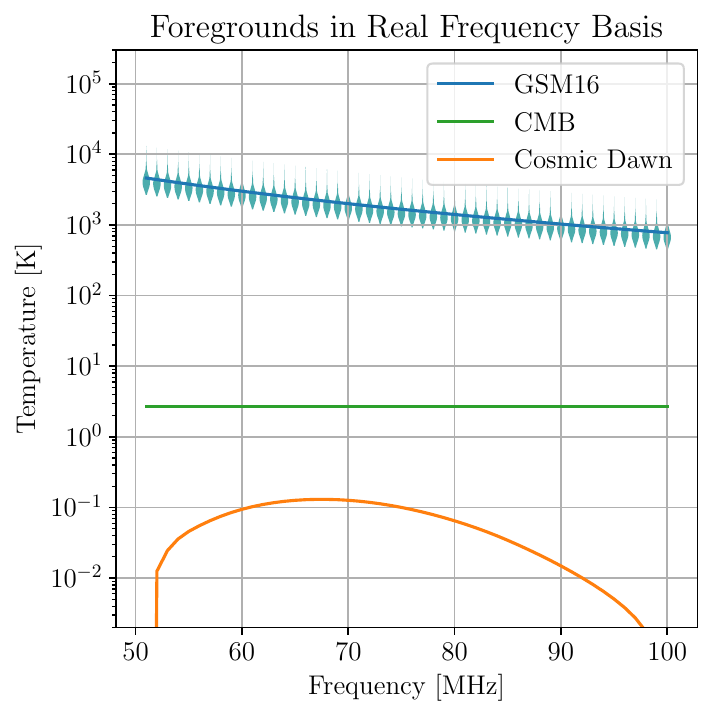}
    \includegraphics[width=0.49\linewidth]{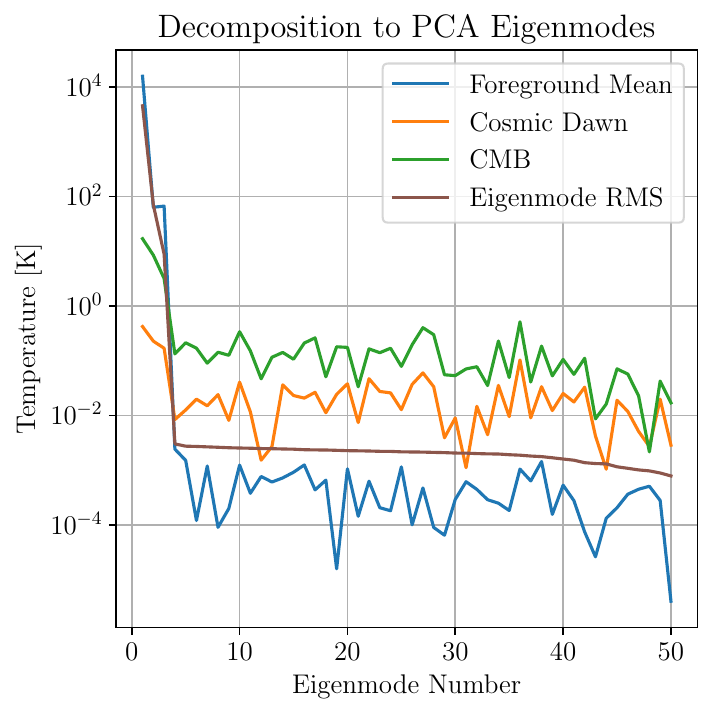}
    \includegraphics[width=1.0\linewidth]{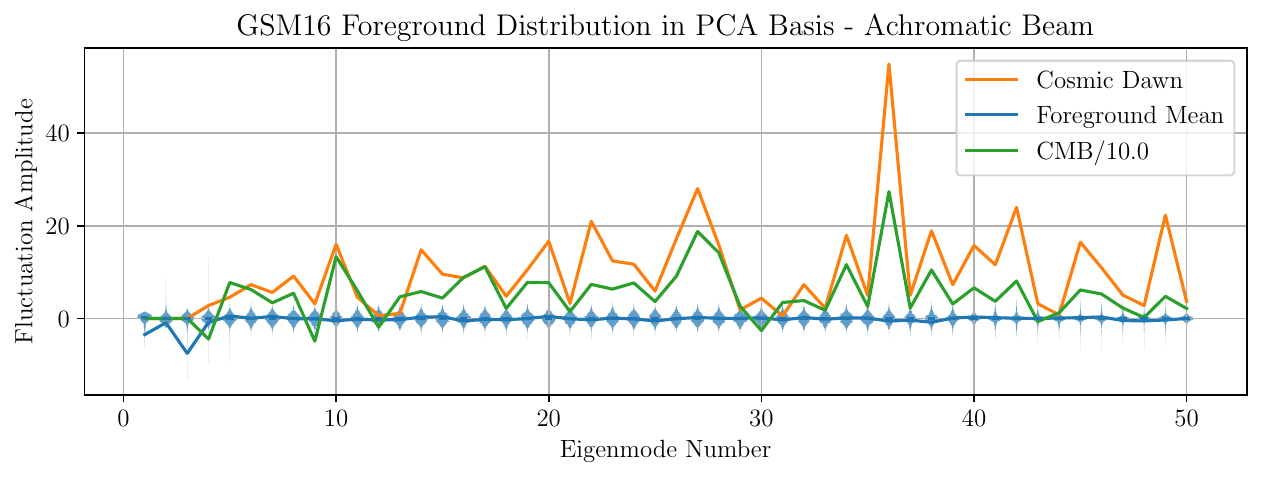}
\caption{This figure shows the separation of foregrounds and the Cosmic Dawn signal in different spaces for the Global Sky Model. \emph{Top Left}: Separation in the original frequency space using an achromatic $2^\circ$ beam. The blue distribution violins and the solid line describe the foreground signal distribution and mean (on the cut sky), the green line denotes the CMB signal, and the orange solid line denotes the Cosmic Dawn signal. Note the extremely large dynamic range on the vertical axis. \emph{Top Right}: The amplitude of the same signals in the foreground PCA basis. The first six principal components absorb about 6 orders of magnitude of the signal. The brown solid line denotes the RMS of the corresponding eigenmode. \emph{Bottom Panel} Signal amplitudes scaled by the RMS of the corresponding eigenmode. The basis vectors have been multiplied by $-1$ to make all the Cosmic Dawn contributions positive.\label{fig:gsmpca}}
\end{figure*}

We start our analysis by taking the foreground fluctuation vector at each point $\DTfg(\vxh) = T(\vxh) - \Tfg$, rotating it into its principal component basis in the frequency space, and normalizing it by dividing by the root-mean-square of variance in each basis direction. 
%via $\mathbf{U} \in \mathbb{R}^{n \times n}$ where the columns of $\mathbf{U}$ are the eigenvectors of $\DT \DT^T$. We further normalize the map by dividing by the RMS of the fluctuations in each frequency bin.
%\begin{equation}
%    \F=\mathbf{U}^T \DT / \rm RMS (\mathbf{U}^T\DT)
%\end{equation}
%This $\F \in \mathbb{R}^{n \times p}$ represents the distribution of foreground fluctuations in the eigenmode basis. 
In other words, for each direction on the sky, we have replaced the description of the data as a vector of binned frequency measurements to a rotated vector in the PCA eigenbasis. The PCA eigenmodes depend on the beam, chromaticity and the galactic cut since they effectively express the eigenshapes of the foregrounds with the most to least amount of variance in fluctuations across the sky. Intuitively, this basis is independent of the spectral shape of both the Dark Ages and Cosmic Dawn signals, which have approximately constant variance across the sky. This makes the lower order PCA eigenmodes with the most foreground variance dominate the signal, compared with the higher order modes where the foreground variance is low.

Figures \ref{fig:pca} and \ref{fig:gsmpca} show how each signal looks in the original frequency space and the corresponding foreground PCA eigenspace. In the top panel, we see the many orders of magnitude that separate the signal of interest from the foreground. The middle panel demonstrates that the first 20 or so eigenmodes absorb the majority of the Dark Ages foreground complexity and that signal of interest does dominate the foregrounds for the sufficiently high-order components. For the Cosmic Dawn case, we see that less than 6 of the first eigenmodes are sufficient to capture most of the foreground variation, allowing the signal to dominate the rest of the eigenspace. Note that this plot suppresses the cross-correlations between the foreground components, which ultimately helps identify the signal of interest as separate from the foreground distribution.

There is no loss of information in this change of basis (rotating into the PCA basis is an invertible, linear transformation), but it makes the subsequent normalizing flow process considerably more numerically robust. We have also found that the naive calculation of principal components from the covariance matrix is numerically unstable due to the extensive dynamic range of data. Therefore, we implement this rotation using the singular vector decomposition approach. Further discussion of this issue can be found in the Appendix \ref{app:pca}.

\subsection{Building $\Pfg$ using Normalizing Flows}

We proceed to model the foreground fluctuations in the PCA space using the Sliced Iterative Normalizing Flow (SINF) implementation of the Normalizing Flow model developed in \cite{dai_sliced_2021}. The algorithm is stable, fast to train, and performs well on small datasets for density estimation and out-of-distribution detection. For a given beam, we subsample the points for efficiency and use a randomly chosen $80\%$ of these points to train the model and the rest for validation.

After training, the model learns several thousand layers of affine transformations $\mathcal{N}=\{\mathcal{N}_i\}:\mathbb{R}^n \rightarrow \mathbb{R}^n$ to map the input distribution to a Gaussian. The log-likelihood for any trial vector $\Tfg$ can then be found as the usual change of variables given by the sum of the Gaussian log-likelihood with the log Jacobian $\mathcal{J}$ of the transformation
\begin{equation}
    \log \Pfg(\Tfg) = \log P(\mathcal{N}(\Tfg)) + \log \mathcal{J}
\end{equation}

The SINF engine has several tunable parameters that regularize the normalizing flow to reduce overfitting and improve performance. We set these parameters to heavily regularize the data to avoid overfitting, which comes at the cost of slow training and a large number of iterations required for convergence. However, this is required since overfitting is an important issue to mitigate in any distribution modelling task. 

The SINF engine is inherently stochastic under the hood. While the results should not be affected, we find a very weak variance across runs on identical training data. This is elaborated in Appendix \ref{app:sinf_seed}

\subsection{Multiscale beams}

In our default implementation, the distribution of foreground fluctuations considers all data randomized with respect to spatial positions. Hence, any spatial correlations in the foregrounds are lost, which is sub-optimal. To learn spatial correlations and scale dependence, if any, we consider stacked maps with beams of different scales, which we call multi-scale maps. 

Given a map with a $2^\circ$ beam (say), we can synthetically smooth it further using a $4^\circ$ and $6^\circ$ beam. We then concatenate the data vectors for every direction from these maps to form a single data vector thrice the original size. This procedure effectively introduces correlations at every direction of $4^\circ$ and $6^\circ$ scales. Note that these do not correspond to the physical beams of the instrument. Instead, they are constructed as further coarse-grained maps of the base map. When considering the multi-scale fits, we first add noise to the base map and then smooth this map further to obtain the rest of the multi-scale maps before concatenating. We are in effect simulating a 2 degree experiment map on which the multi-scale technique has been applied.

We then calculate the PCA basis for this multi-scale map and model the fluctuations in this multi-scale map as before. In this preliminary work, we consider 4$^\circ$ and 6$^\circ$ degree beams in addition to the base 2$^\circ$ beam. The most robust way of constructing the multi-scale map is to concatenate only the differences with the base map; that is, instead of the naive $2^\circ + 4^\circ + 6^\circ$ multi-scale map, we use $2^\circ + (4^\circ-2^\circ) + (6^\circ-4^\circ)$. We believe this reflects exactly the extra information contained in spatial correlations at $4^\circ$ and $6^\circ$ scales for any direction and also ends up being numerically stable. For brevity, we will still refer to these as $2^\circ+4^\circ$ and $2^\circ+4^\circ+6^\circ$ beams. Since the monopole signals are perfectly spatially correlated (and hence independent of beam scale), the multi-scale version for the monopole signals is given by the monopole signal concatenated with zeroes.

We have tested the impact of filtering PCA components and found that while there was some variation in the results, there were no clear advantages or disadvantages in doing so.

\subsection{CMB Signal Model}

To test our method, we have considered a model with free CMB temperature. This model is trivial; it is just a frequency-independent constant $T_{\rm CMB}(\nu) = 2.718$. This parameterization allows us to ask just how sensitive a putative futuristic experiment would be to the CMB temperature deep in the Rayleigh-Jeans limit. At the same time, it would allow us to tie the instrument's calibration against a very robust CMB as a baseline. 

The constant CMB temperature in frequency space translates to a non-trivial vector in the foreground PCA basis, which, in comparison with the Dark Ages and Cosmic Dawn signals, is a much stronger monopole signal, as can be seen in Figures \ref{fig:pca} and \ref{fig:gsmpca}.

\begin{figure}
    \centering
    \includegraphics[width=\linewidth]{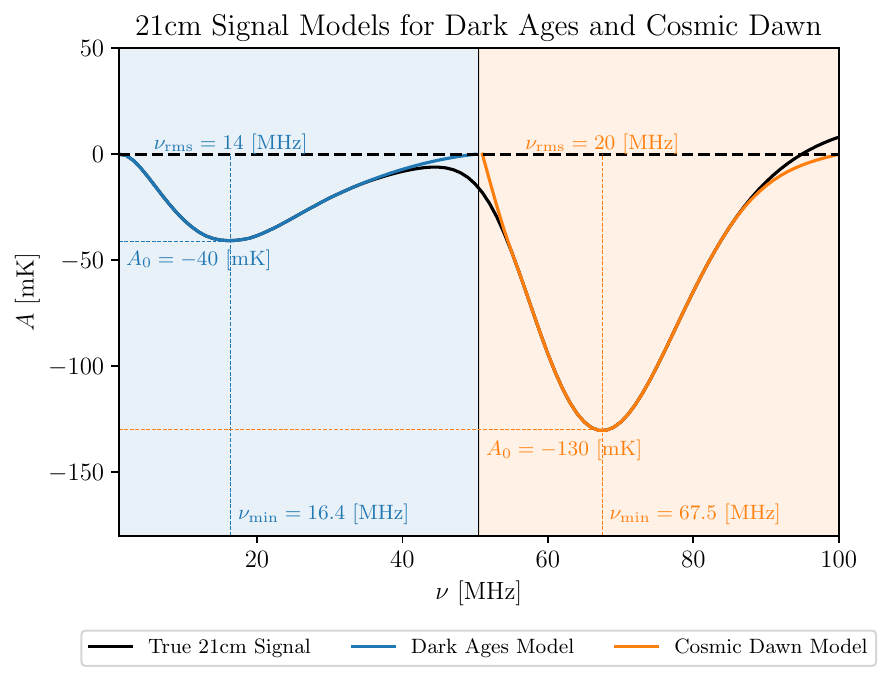}
    \caption{Model approximations to the fiducial 21cm signal used in this analysis. Both the Dark Ages and Cosmic Dawn models are 3-parameter models that depend on the amplitude, width and center frequency, and are scaled to approach zero at the boundaries of their domains.
    \label{fig:models}}
\end{figure}

\subsection{21\,cm Signal Models}

We build phenomenological 3-parameter models for the Dark Ages and Cosmic Dawn signals as follows. We start with the fiducial 21\,cm signal shape as predicted by the ARES software package by \cite{mirocha_decoding_2014}. We then scale this model in amplitude, position, and width:

\begin{equation}
    T_{\rm 21} (\nu, A, \nu_c, \Delta \nu) = \frac{A}{A_0} T_{\rm ARES}\left( \nu_{c0}+(\nu-\nu_c)\frac{\Delta \nu}{\Delta \nu_{0}} \right)
\end{equation}
where $A_0=40\,$mK, $\nu_{c0}=16.4\,$Mhz and $\Delta \nu_{0}=14\,$Mhz are (defined to be) values of amplitude, trough center and width for the Dark Ages signal and $A_0=130\,$mK, $\nu_{c0}=67.5\,$Mhz and $\Delta \nu_{0}=20\,$Mhz are (defined to be) the values of amplitude, trough center and width for the Cosmic Dawn Signal. To avoid the method latching on to the mid-frequency inflection region at 50\,MHz, we have manually modified both the models so that they gracefully approach zero at 50MHz by matching first and second derivatives at 32\,MHz, 55\,MHz and 85\,MHz. See Figure \ref{fig:models} for the exact model shapes compared to the fiducial 21cm signal.

\subsection{Likelihood analysis}

It is clear from Figures \ref{fig:pca} and \ref{fig:gsmpca} that some linear combinations of both the Dark Ages and Cosmic Dawn signals are subdominant to the foregrounds, thus making it difficult to recover the full unbiased signal template. We may, however, calculate the relative likelihoods of one signal template to another. 

Our signal model has three degrees of freedom, and therefore, the likelihood analysis can be done by brute-force evaluation on a sufficiently fine 3D grid ($N=70^3$). In both the 3D grid and the 1D amplitude likelihoods, we use flat priors for the respective parameter-spaces. For the Dark Ages 3D grids, we use the priors $A\in(0.4\,{\rm mK},400\,{\rm mK}), \nu_{\rm rms}\in(10\,{\rm MHz},30\,{\rm MHz}), \nu_{\rm min}\in(10\,{\rm MHz},30\,{\rm MHz})$, and for Cosmic Dawn we use $A\in(1.3\,{\rm mK},1300\,{\rm mK}), \nu_{\rm rms}\in(10\,{\rm MHz},40\,{\rm MHz}), \nu_{\rm min}\in(50\,{\rm MHz},90\,{\rm MHz})$ for the signal amplitude, width and center respectively. For Dark Ages, we initially tested the 3D grid limits by imposing the physical $\nu_{\rm rms},\nu_{\rm min}\in(1\,{\rm MHz},50\,{\rm MHz})$ for the both the signal width and center, which correspond to the bin width and maximum frequency which was further reduced to the limits above to improve the grid density around the regions of interest while leaving the likelihoods unchanged. For Cosmic Dawn, we similarly tested with the full $A\in(50\,{\rm MHz},100\,{\rm MHz})$ range for center, $\nu_{\rm rms}\in(10\,{\rm MHz},40\,{\rm MHz})$ range for width, and $\nu_{\rm min}\in(50\,{\rm MHz},90\,{\rm MHz})$, and reduced it to the above. For the 1D amplitude likelihoods we use the flat priors $(0.1A_0,1000A_0)$ where $A_0=40$\,mK for Dark Ages and $130$\,mK for Cosmic Dawn, respectively.

After we evaluate likelihood on a regular grid, we use the standard python MCMC visualization package \texttt{corner} from \cite{corner} to plot and calculate marginalized constraints by assigning sample weights proportional to their likelihood. Where the likelihood is prior-dominated, we instead calculate the amplitude-only likelihoods by fixing the width and center to their fiducial values, giving us at least the exclusion limits on the amplitude of the signal of interest. 

We also analyze the effects of independent noise realizations on the likelihood constraints. We generate ten independent noise realizations for each SNR level and calculate the likelihood constraints for each realization. We then calculate the median likelihoods and variation across the different noise realizations.

\section{Results}
\label{sec:results}

\begin{figure*}
    \centering
    \includegraphics[width=\textwidth]{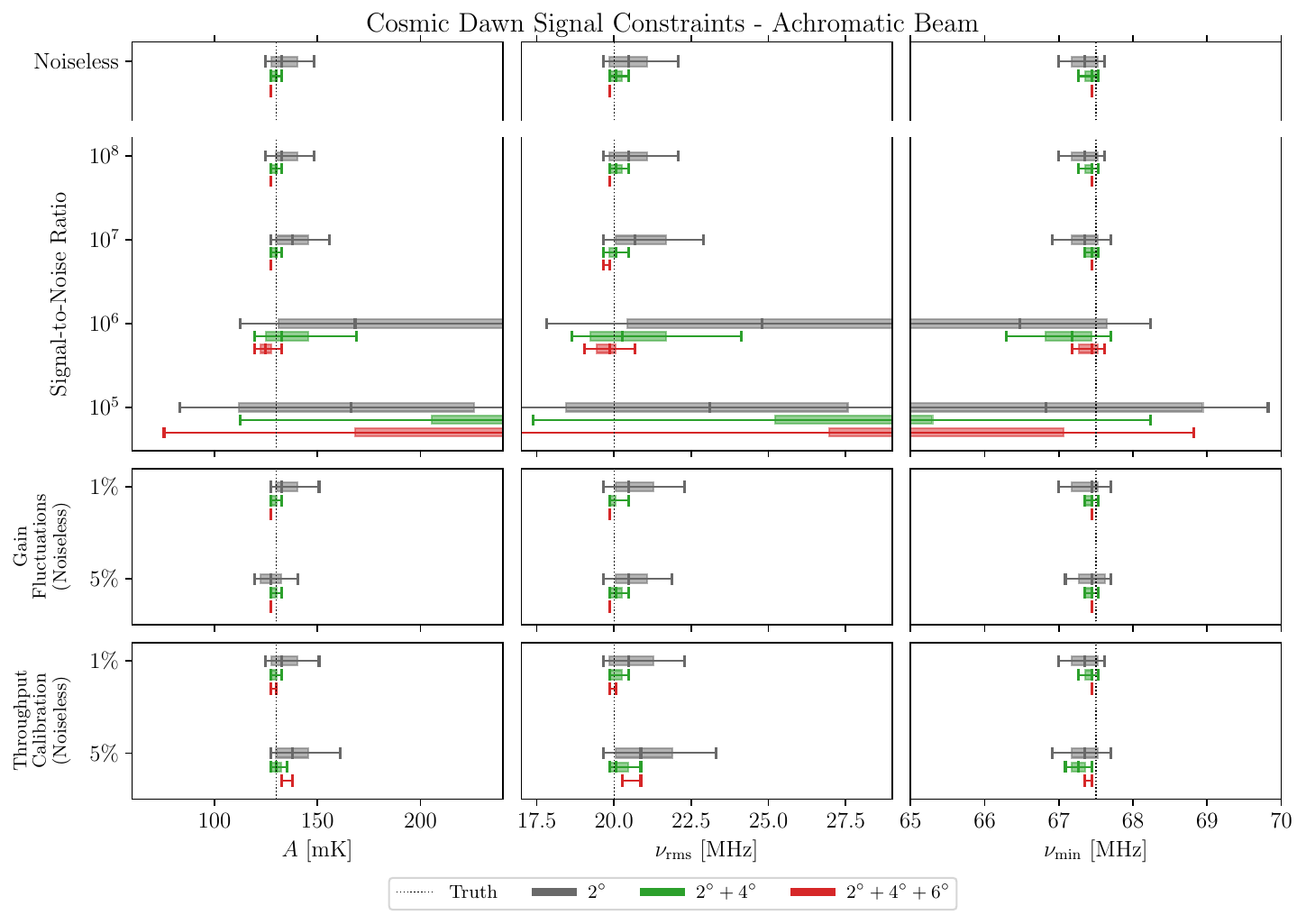}
    \caption{Cosmic Dawn signal constraints for achromatic beams for the ideal noiseless, noisy, and noiseless instrumental systematics cases. Box plot boxes represent 1-sigma errors, median line in the center, and whiskers bound the 95\% confidence intervals for each case. Gray areas represent the physical prior on the three parameters and dashed black indicate the fiducial values of the parameters. All results correspond to the same fixed noise realization. 
    \label{fig:CD_achromtable}}
\end{figure*}

\begin{figure*}
    \centering
    \includegraphics[width=\textwidth]{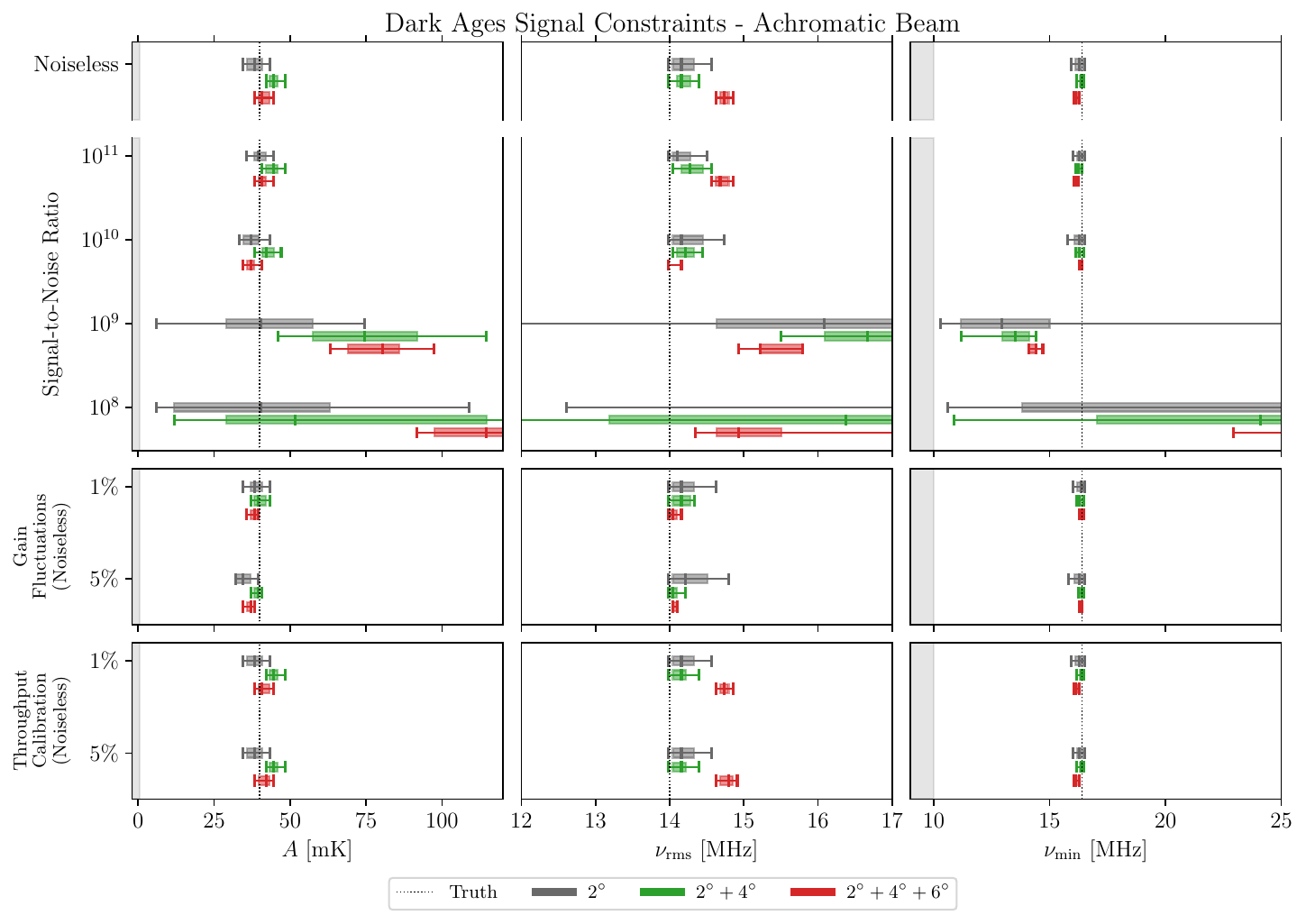}
    \caption{Dark Ages signal constraints for achromatic beams for the ideal noiseless, noisy, and noiseless instrumental systematics cases. Box plot boxes represent 1-sigma errors, median line in the center, and whiskers bound the 95\% confidence intervals for each case. Gray areas represent the physical prior on the three parameters and dashed black indicate the fiducial values of the parameters. All results correspond to the same fixed noise realization. 
    \label{fig:DA_achromtable}}
\end{figure*}

\begin{figure*}
    \centering
    \includegraphics[width=\textwidth]{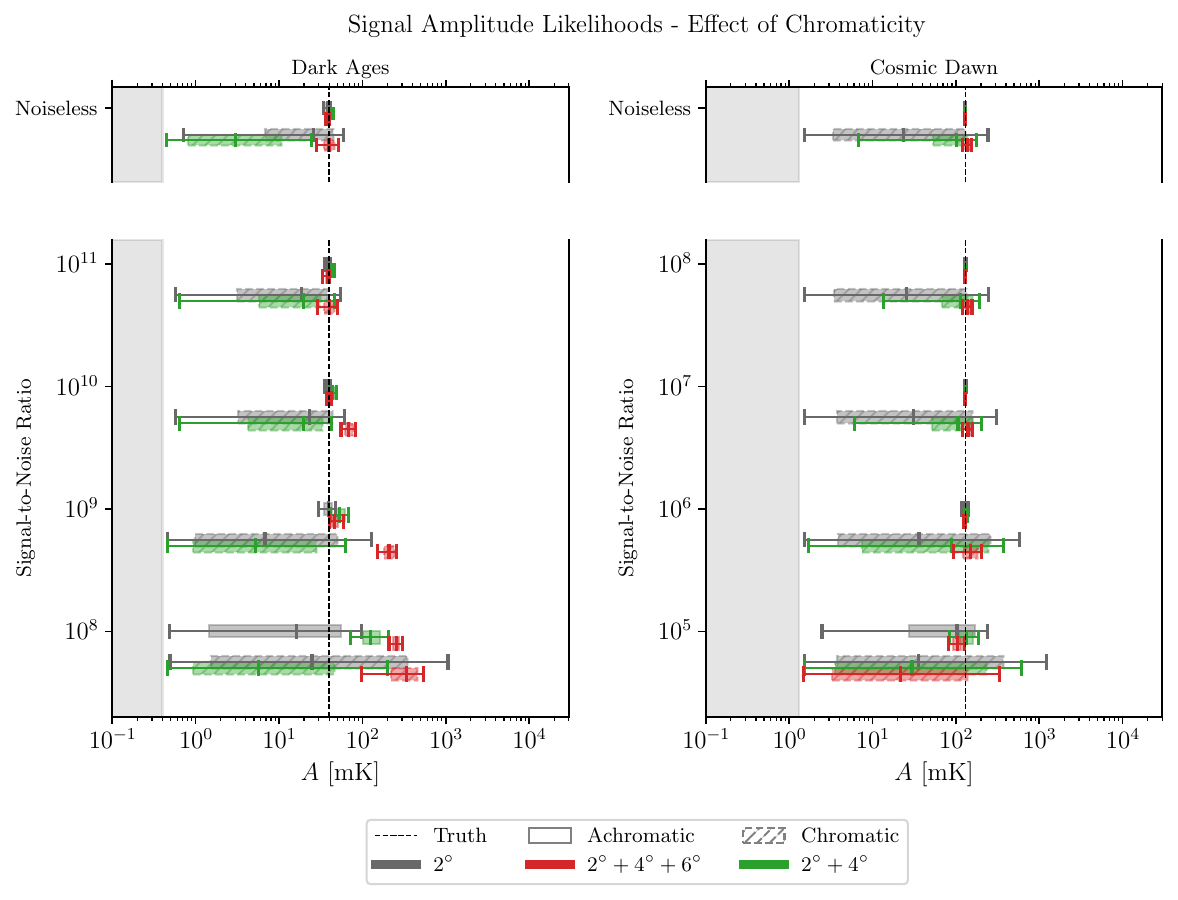}
    \caption{Signal amplitude likelihood constraints for achromatic and chromatic beams, with the width and center fixed to the fiducial values, for the ideal noiseless, noisy, and noiseless instrumental systematics cases. Box plot boxes represent 1-sigma errors, median line in the center, and whiskers bound the 95\% confidence intervals for each case. We plot achromatic (top pair, regular) and chromatic beams (bottom pair, hatched) for each combination. Gray areas represent the physical prior on the three parameters and dashed black lines indicate the fiducial values of the parameters. All results correspond to the same fixed noise realization. \label{fig:chromaticity}}
\end{figure*}

In Figures \ref{fig:CD_achromtable}, \ref{fig:DA_achromtable} and \ref{fig:chromaticity}, we show the likelihood constraint results for the various tests that we have performed on both the 21cm signal models. When the signal-to-noise is sufficiently high, we are able to recover all three of the signal parameters. At low signal-to-noise, the 3-parameter marginalized likelihoods are prior-dominated and highly non-Gaussian. In the case of chromatic beams, we show the 1D amplitude likelihoods since we find that the full 3-parameter marginalized likelihoods are consistent with zero amplitude, leaving the width and center completely unconstrained. 

\begin{figure}
    \centering
    \includegraphics[width=0.49\textwidth]{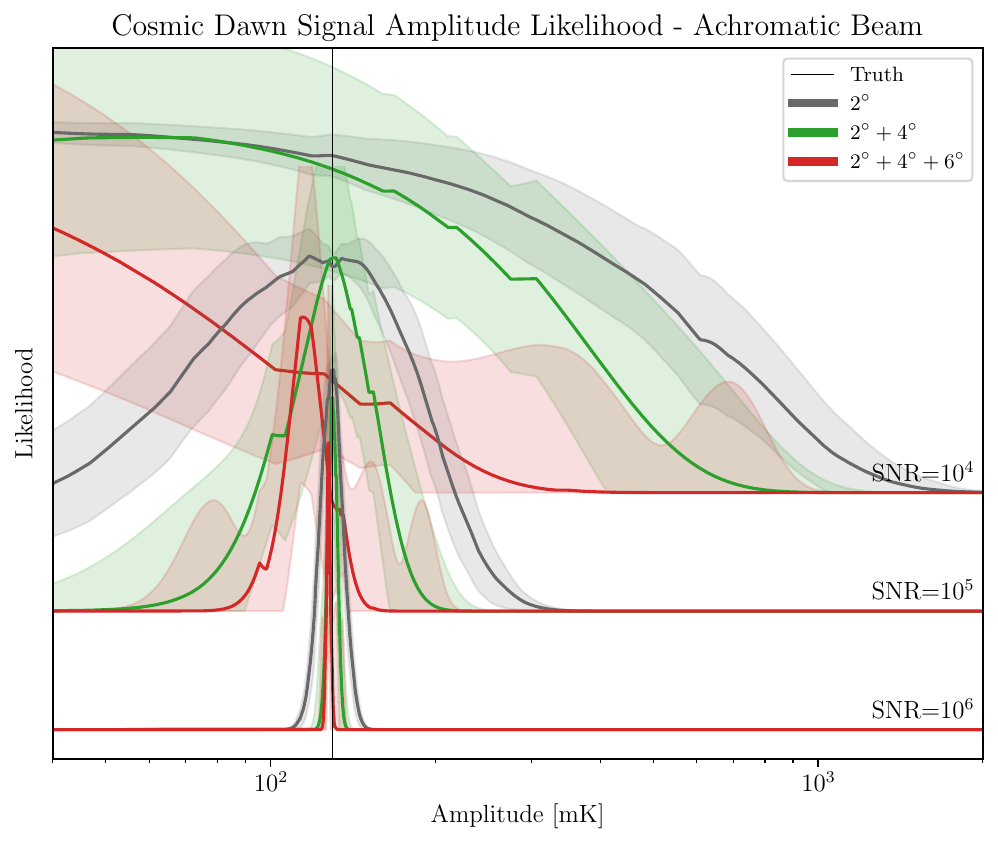}
    \includegraphics[width=0.49\textwidth]{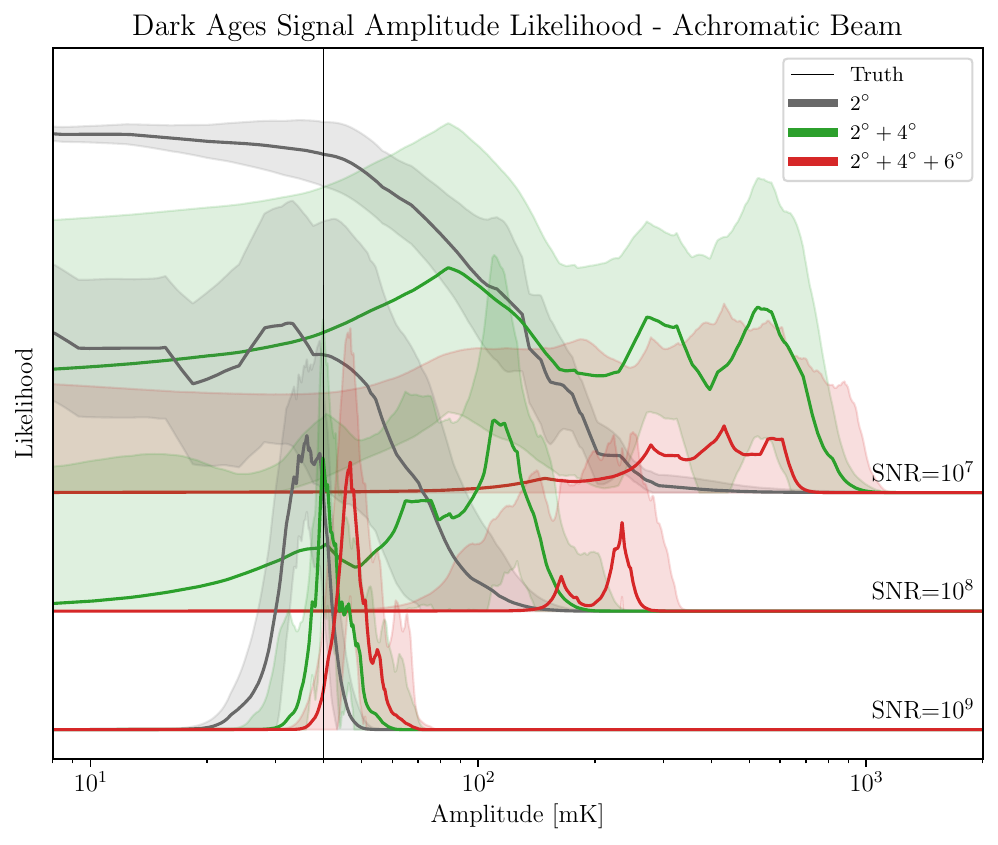}
     \caption{Signal amplitude likelihoods for the Cosmic Dawn signal and the Dark Ages signal for ten different noise realizations at different signal-to-noise levels for achromatic beams. The solid lines represent the median likelihoods over all noise realizations, and the shaded regions correspond to the 1-sigma variation across the different noise realizations, with an added offset to aid readability. The vertical line at $40$\,mK and 130\,mK is the fiducial value for the signal amplitude for the Dark Ages model and Cosmic Dawn model, respectively.
     \label{fig:Amp1D24}}
\end{figure}

\begin{figure}
    \centering
    \includegraphics[width=0.49\textwidth]{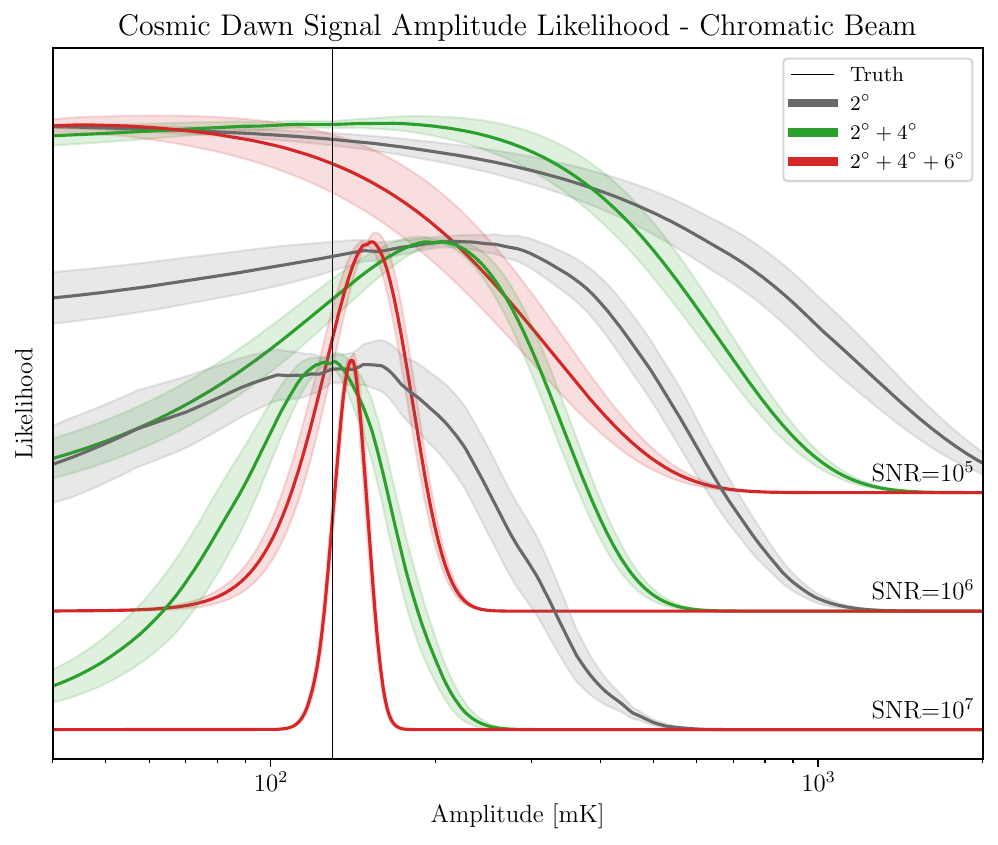}
    \includegraphics[width=0.49\textwidth]{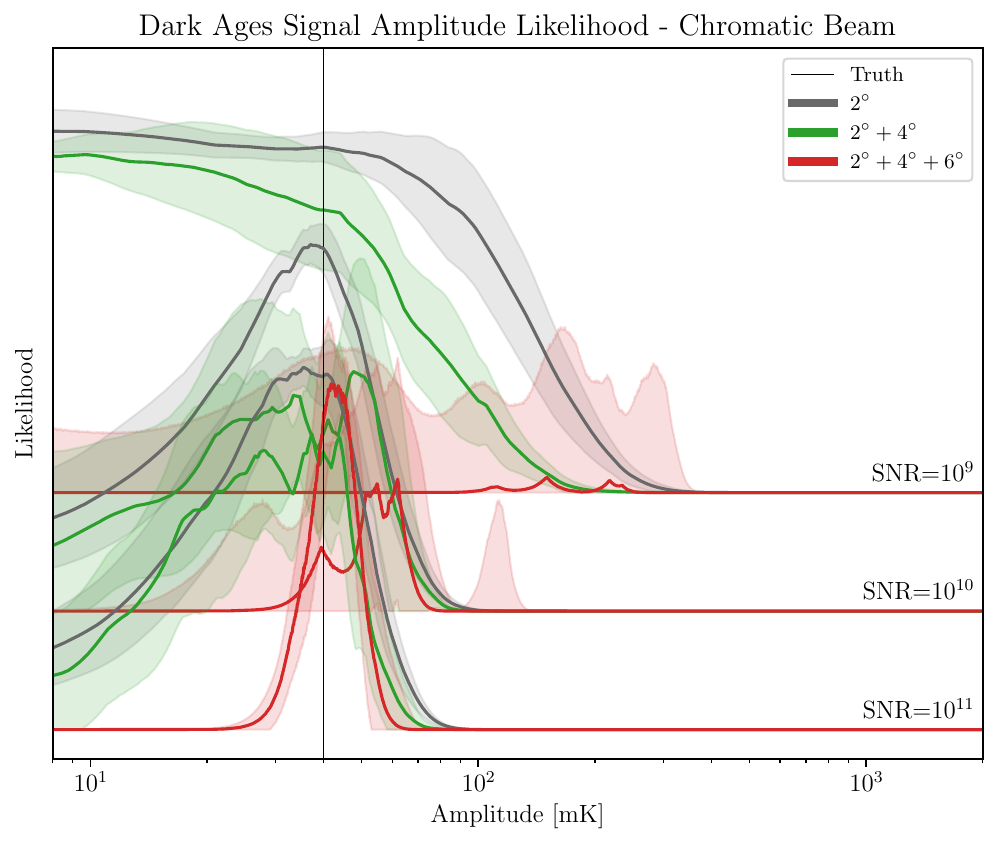}
    \caption{Signal amplitude likelihoods for the Cosmic Dawn signal and the Dark Ages signal for ten different noise realizations at different signal-to-noise levels for chromatic beams. The solid lines represent the median likelihoods over all noise realizations, and the shaded regions correspond to the 1-sigma variation across the different noise realizations, with an added offset to aid readability. The vertical line at $40$\,mK and 130\,mK is the fiducial value for the signal amplitude for the Dark Ages model and Cosmic Dawn model, respectively. 
    \label{fig:Amp1Dc}}
\end{figure}

Figures \ref{fig:Amp1D24} and \ref{fig:Amp1Dc} show the 1D amplitude likelihoods for the Dark Ages and Cosmic Dawn signals for ten different noise realizations at different signal-to-noise levels. Across noise realizations, we find that multi-scale fits seem to display realization-dependent biases in the amplitude constraints. 

\subsection{Results in an idealistic case}

Our idealistic case is a perfectly calibrated, noiseless instrument. Results from the Figures \ref{fig:CD_achromtable}, \ref{fig:DA_achromtable}, and \ref{fig:chromaticity} can be summarized as follows. First, we find that in majority of cases we have detection of the signal with high confidence. The chromatic multi-scale case recovers the fiducial values of parameters with percent level confidence. We also find that the results are unbiased, with the exception of $2^\circ+4^\circ+6^\circ$ achromatic Dark Ages case (discussed below), the results are consistent with the ground truth at two sigma level, even when sensitivity is at the percent level.

This is an important result that explicitly demonstrates that, in principle, the foregrounds are separable from the signal of interest for a sufficiently sensitive instrument. Even though these maps are noiseless, there is a residual uncertainty in the amplitude reconstruction given by the foreground complexity at the percent level in amplitude. If the real foregrounds are more complex than those in the GSM16/ULSA maps (as is almost certainly the case), this uncertainty would grow somewhat. Conversely, if the signal level is not what we expect, our ability to measure it would change. For example, were the real signal 100 times weaker, then no instrument would be capable of performing foreground separation, even at the foreground complexity present in the GSM and ULSA maps\footnote{Of course, it is plausible that some new clever statistical method might be invented.}  

Our second finding is that multi-scale fits help a lot in improving constraints. Especially in the chromatic cases, combining three scales improves the uncertainty on the amplitude parameter by a factor of $\sim 10$ over the single scale. Similar improvements are also seen for achromatic cases where all three parameters of the 21cm signal models see improved constraints. Multi-scale maps bring in information about spatial correlations from different scales which is crucial to mitigate the effects of chromaticity, which is essentially coupling information from different scales with frequency bins.

Finally, beam chromaticity seems to be a real issue because it makes measured foregrounds considerably more complex, as has been discussed in the literature, for example, \cite{2011.00549, ansteyGeneralBayesianFramework2021,saxenaSkyaveraged21cmSignal2023}, etc. For individual single-scale measurements, the errors increase by up to a factor of 100, making it only possible to set an upper bound on the signal amplitude. Multi-scale fits can remedy this significantly for the 1D amplitude likelihoods, and we illustrate this in Figure \ref{fig:chromaticity}. The full 3D likelihoods are still prior-dominated and do not provide any useful constraints.

\subsection{Effects of adding noise}

We now consider the effect of adding noise to the maps. We find that for the pessimistic noise scenario (see \ref{tab:snr_scenarios}), with an SNR of 10$^8$ for the Dark Ages signal and 10$^5$ for the Cosmic Dawn signal, it is possible to constrain the upper limit on the amplitude of the signal of interest with some confidence, especially in the achromatic case (see Figures \ref{fig:chromaticity}, \ref{fig:Amp1D24}, \ref{fig:Amp1Dc}). For chromatic beams, the amplitude likelihood constraints for Dark Ages multi-scale fits are biased towards higher values and inconsistent with the fiducial amplitude. We find that independent noise realizations can lead to different values of this bias. However, the sizes of the errors remain consistent (see Appendix \ref{app:sinf_seed} for details on systematic and noise-realization biases). The three-parameter likelihoods are prior-dominated and do not provide any useful constraints at this SNR level.

In the realistic signal-to-noise scenario of SNR $10^9$ for Dark Ages and $10^6$ for Cosmic Dawn, we find that the foreground models begin to be able to constrain the three signal parameters well for both Dark Ages and Cosmic Dawn, with achromatic beams and especially when using multi-scale fits. Multi-scale $2^\circ+4^\circ+6^\circ$ fits also enable constraining the chromatic case for both models since both the $2^\circ$ and $2^\circ+4^\circ$ fits are only able to put an upper bound on the signal amplitudes. The results are consistent with the idealistic case, with the exception of the Dark Ages signal amplitude in the chromatic case, which is biased towards higher values in this case.

For the optimistic scenario of SNR $10^{10}$ for Dark Ages and $10^7$ for Cosmic Dawn, we find that the foreground models are able to constrain the three signal parameters to percent-level precision, for both Dark Ages and Cosmic Dawn, for achromatic beams and when using multi-scale fits. The results are consistent with the idealistic case, with the exception of the Dark Ages signal amplitude in the chromatic case, which is biased towards higher values in this case.

The results are further scrutinized in Figures \ref{fig:Amp1D24} and \ref{fig:Amp1Dc}, where we look at the amplitude likelihood constraints for several different noise realizations. We find that averaged over multiple likelihoods, the correct amplitude of the Dark Ages signal is measured, and that results may have a bias depending on the noise realization. For more details on the noise realizations, see Appendix \ref{app:sinf_seed}.

\subsection{Instrumental effects}

We study two instrumental effects: gain fluctuations and overall bandpass structure.

\subsubsection{Temporal Gain fluctuations}
\label{sec:gain}
Gain fluctuations are inevitable in any radiometer using contemporary technology. Over time, we may expect the total system gain to fluctuate, leading to some directions in the sky being measured with different amplitudes, and there will be residual effects even after calibration. In most instruments, the dominant effect is the frequency-independent change in amplitude. In practice, there are subdominant temporal changes in spectra response, but we ignore these in this paper.
In standard smoothness-based foreground separation approaches, they are completely harmless because they do not change the smoothness of either foregrounds or the signal of interest. In our case, however, they are of potential concern because they leak the signal of interest into signal fluctuations. Concretely, for a sky direction for which the frequency  gain $g$ has drifted from unity, we have
\begin{equation}
    \DTfg \rightarrow g \DTfg + (1-g) \bar{T}_{\rm fg} + (1-g) \Ts.
\end{equation} 

While the first two terms are likely harmless, the last term will leak the shape of the signal into fluctuations. The foreground model will then recognize this shape as foreground at the level of $(1-g)$. We, therefore, expect the ability to measure the signal amplitude to be limited to the RMS of gain fluctuations. 

Gain fluctuations do average down at least partly over time and number of observations. If the same patch of the sky is observed $N$ times, the RMS of gain fluctuations will decrease by $1/\sqrt{N}$. In order to simulate the most realistic $1/f$ type gain fluctuation effects, we generate a Gaussian random field smoothed the same as the base observation and then generate gain fluctuations consistent with such level. Results are also in Figures \ref{fig:DA_achromtable} and \ref{fig:CD_achromtable}. We see that if gain fluctuations do hamper our ability to extract the signal, they do so modestly. 

\subsubsection{Overall throughput calibration}

Smoothness-based separation is based on the premise that foregrounds change more gradually across frequency space compared to the 21cm signal. Consequently, any systematic effects that induce losses or gains in the signal pathway from the antenna to the receiver can significantly influence the outcome. A typical example is cable reflection, which introduces a sinusoidal pattern into the throughput's overall profile. If these effects are known and measured they can be calibrated out, but often they are not or there are uncalibrated residuals. To study robustness to this type of error we consider a simple multiplicative error that is a function of frequency, but independent of time or position on the sky.

In our procedure, any spectral distortion function $g(\nu)$ that is purely multiplicative in frequency only changes the eigenvectors of the problem, which also get multiplied by $g(\nu)$. Everything else, including the decomposition into the PCA basis and the likelihood calculation, remains the same. So, ultimately, the primary error we are making is using the wrong basis vectors when calculating the likelihood of a given theoretical model. So the net effect of this type of error is that we distort the model we are studying by the unknown systematic.

To calculate the impact of this systematic, we consider a ``cable reflection'' type effect with 
\begin{equation}
    g(\nu) \rightarrow \left( 1+ A_c \cos (\omega_c \nu) \right),
\end{equation}
where we fix $\omega_c$ to have three full cycles across the 50MHz band and vary $A_c$.

Results are also shown in Figures \ref{fig:DA_achromtable} and \ref{fig:CD_achromtable}. Note that we do not expect a change in accuracy in this case, but we might expect a bias in the results. While we see that results do change, they are surprisingly stable. Using an incorrect basis seems to affect the multi-scale fits more than the single-scale fits, where we see a bias in the constraints when comparing the 5\% and 1\% cases.

\subsection{Detecting CMB}

\begin{figure*}
    \centering
    \includegraphics[width=\textwidth]{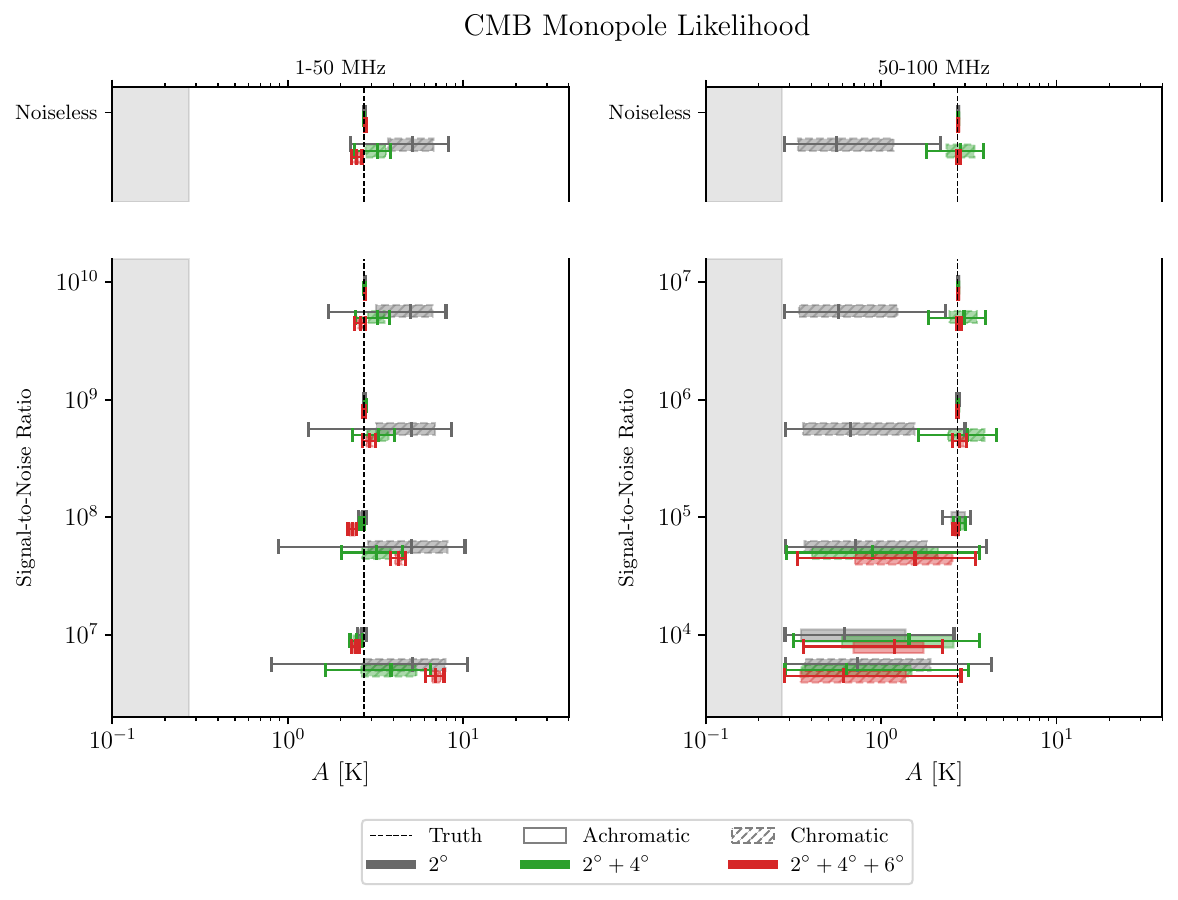}
    \caption{CMB Signal constraints for achromatic(filled box plots) and chromatic (hatched box plots) beams for the ideal noiseless and noisy cases. Box plot boxes represent 1-sigma errors, median line in the center, and whiskers bound the 95\% confidence intervals for each case. Dashed black indicates the fiducial values of the parameters. All results correspond to the same fixed noise realization.
    \label{fig:CMB_table}}
\end{figure*}

\begin{figure*}
    \includegraphics[width=0.49\textwidth]{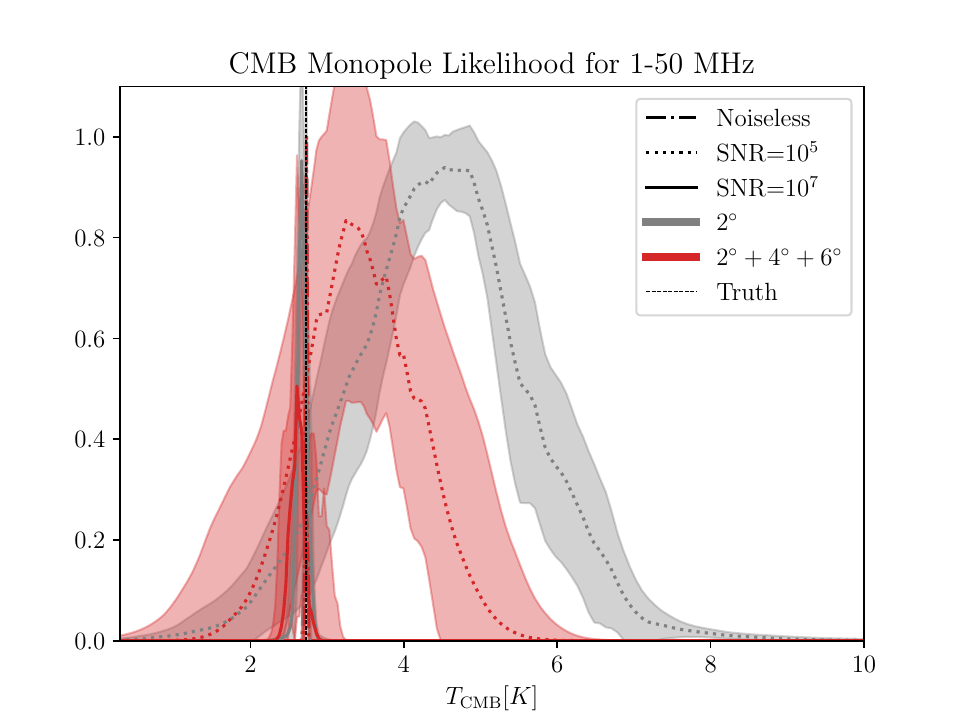}
    \includegraphics[width=0.49\textwidth]{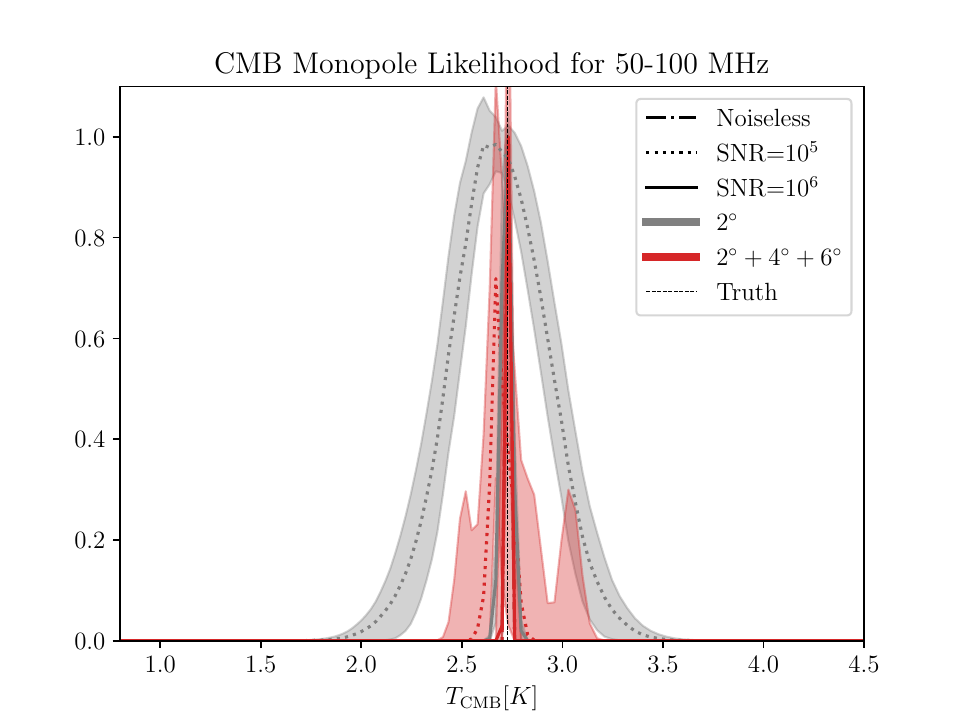}
    \caption{\label{fig:AmpCMB} CMB amplitude constraints for an achromatic $2^\circ$ and $2^\circ+4^\circ+6^\circ$ beams at various noise levels. Central lines represent the median likelihoods, and shaded regions represent the error bounds derived from a set of 10 different noise realizations.}
\end{figure*}

In Figure \ref{fig:CMB_table}, we present the results for the CMB monopole signal for a $2^\circ$ instrument with and without chromaticity, along with the improvements due to a multi-scale $2^\circ+4^\circ$ and $2^\circ+4^\circ+6^\circ$ fits. We see a surprisingly strong detection of the signal. While CMB is very smooth, it simultaneously has a very different spectral shape, being a power-law with spectral index zero compared to foregrounds, which are a power-law with a spectral index $\approx -2.5$. CMB signal is recovered very well at moderate noise levels as one would expect. The CMB signal, being two orders of magnitude large in amplitude compared to the Dark Ages signal, should be the prime target for the first generation of Dark Ages experiments. 

Figure \ref{fig:AmpCMB} shows some of the typical likelihood surfaces associated with the CMB temperature.

\section{Discussion \& Conclusions}
\label{sec:conclusions}

We have presented a new method for extracting the presence of a monopole signal in radio experiments. The method can successfully extract any monopole that is sufficiently statistically different from the foreground signal, assuming the fluctuations in the foreground are representative of their mean signal. 

We have argued that the traditional smoothness-based separation is challenging to put into practice in the presence of strong foregrounds and antenna chromaticities since any residual effects would have to be controlled at a one-part in a million, many orders of magnitude beyond the current state of the art. Instead, we have presented a method where we can build a model for foreground fluctuations in an unsupervised way and apply the model to filter out the foreground monopole. The assumption that this is a valid procedure is based on the physical intuition that the foreground-generating process is local in space. 

We have considered extracting information from multiple scales. We find that fitting multiple scales at once can bring significant benefits in terms of signal-to-noise.

We have demonstrated that the method works for semi-realistic sky models and instrument, but requires high signal-to-noise to really constrain the signal. In a traditional, smoothness-based separation, the requirement is that the noise level is small compared to the signal over the entire sky. In this case, we need to learn about possible spectra shapes of foregrounds \emph{at the level of the signal of interest}, and therefore, we require the noise level to be small compared to the signal in individual resolution elements. Nevertheless, a sufficiently aggressive instrument would reach the necessary sensitivities. For Cosmic Dawn, the required sensitivities are modest by modern standards, and experiments will almost certainly be limited by the systematic effects. But even for Dark Ages, a pessimistic signal-to-noise of $10^8$ can be achieved by $10^3$ independent beams integrating for 12 lunar nights. This is something that is possible within the currently proposed lunar arrays such as ALO \citep{kleinwoltAstronomicalLunarObservatory2024}. We stress that these results are obtained without any specific optimization. By focusing on the quiet parts of the sky with optimized beam sizes, etc., one might be able to do considerably better. 

We have found that while the method is in general unbiased, some of our results are biased, especially when adding the largest 6 degree beams. The natural explanation for this is that at 6 degree smoothing, the number of independent "samples" of the sky is too low to learn the possible shapes. A more in-depth study of when and how this method is robust is left for future work, but we note that biases are generally small compared to the signal.

We found that antenna design matters and that achromatic experiments do significantly better since there is less mixing of foreground shapes. But our method is automatic and does not impose any hard requirements: the less chromatic your instrument response, the better results you will obtain in a self-consistent and self-calibrating manner. Our approach also relaxes requirements on the system throughput shape and the presence of overall gain fluctuations. Perturbations at the 1\% level are sufficient to extract the signal.

These considerations have deep implications for the design of the experiments to measure the 21\,cm monopole signal. Instead of having a low-resolution instrument with an exquisitely calibrated and designed beam, a better strategy is a higher-resolution instrument with small gain fluctuations and a sub-percent level understanding of ground coupling. Such an instrument would target a smaller, quieter patch of the sky to the extreme depths required to measure the signal. The foreground modelling over at high resolution is likely easier, since the emission is more localized, and we have more samples of the foreground fluctuations, enabling us to build a more robust foreground model.

We have also found that our method is surprisingly good at extracting the CMB monopole signal. Measuring the CMB signal deep in the Rayleigh-Jeans regime would be an important milestone for detecting more interesting signals. Assuming alternative scenarios are excluded, this could also act as a calibration signal, tying the overall temperature scale to the precisely known CMB temperature. One might worry that the CMB signal would be perfectly degenerate with the amplifier noise, but this should not be the case in practice. This is because the presence of the ground (or the lunar regolith), as well as the frequency-dependent preamplifier coupling, will make the amplifier noise contribution non-white once the signals are expressed in the units of sky monopole.

There are several avenues for improving and extending our method. The first is to optimize the linear combinations of the raw data that are used for analysis. For example, we have shown that achromatic beams perform better. With a sufficiently high native resolution, it should be possible to generate synthetic achromatic beams that might perform better than raw data. We have found that the particular combination of $2^\circ$, $4^\circ$ and $6^\circ$ beams performs considerably better than individual resolutions without any attempts to fine-tune the results. Still, at the same time, the addition of  $6^\circ$ beams can lead to biases. Understanding where these biases come from and how to avoid (or at least diagnose them) is also of very high importance. In practice,  design and analysis choices can likely be improved, but they are also highly likely to be strongly dependent on the details and complexity of the assumed sky model. Further optimization lies also in the frequency resolution and sky coverage. 

\begin{appendix}

\section{Numerical Care with PCA decomposition}
\label{app:pca}

\newcommand{\X}{\mathbf{X}}
\newcommand{\U}{\mathbf{U}}
\newcommand{\V}{\mathbf{V}}
Typically, PCA decomposition of a data vector $\X\in\mathbf{R}^{m\times n}$ can be done in two equivalent ways viz. eigen-decomposition of the covariance matrix of the data $\mathrm{cov}(\X)=\U S \U^{T}$ where $\mathrm{cov}(\X)=\frac{1}{\sqrt{n-1}}\X^{T}\X$, or singular value decomposition $\U S \V^T = \X$ of the data directly. It is straightforward to show that these methods are identical, up to signs, for calculating $\U$ and their eigenvalues $S$. 

A typical \texttt{float64} number has $\sim15$ digits of precision. If the relative difference between the smallest and largest numbers in your data $X$ exceed $10^8$ in arbitrary units, then squaring the data vector is numerically unstable since the \texttt{float64} data type is unable to offer the required precision for the smallest and the largest entries simultaneously. This is a known numerical issue involving the calculation of the covariance matrix of a data vector that contains elements $O(1)$ along with elements $\epsilon\sim10^{-8} \ll O(1)$. The computation $\X^T\X$ can lead to a loss of precision and give numerically unstable eigenvectors/values. 

\begin{figure}[h!]
    \centering
    \includegraphics[width=0.5\linewidth]{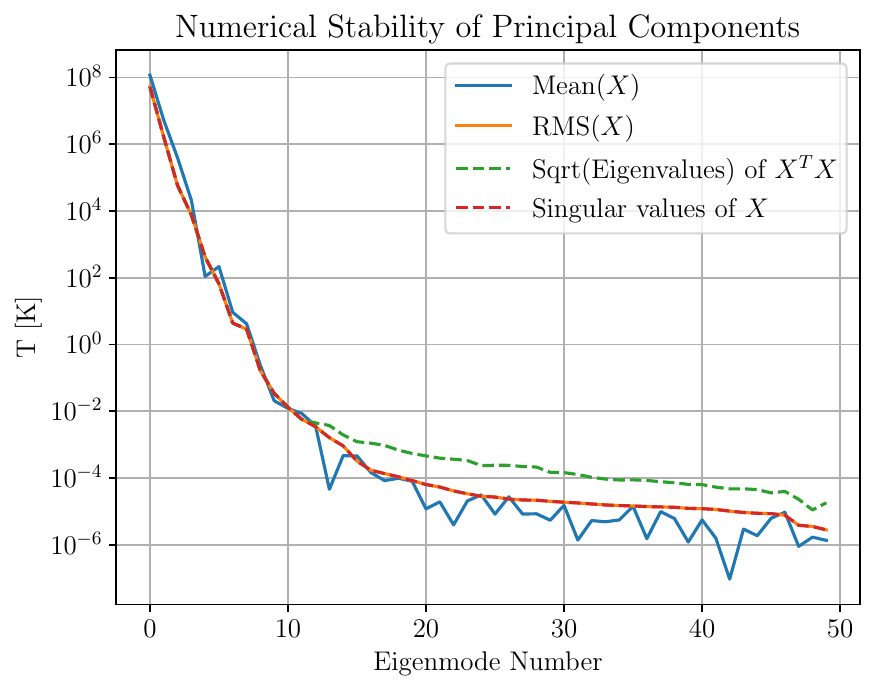}    
    \caption{Numerical stability of the method used for finding the principal components of a data vector $X$. Note the orange line, representing the true RMS values of the data, agrees well with the singular values (dashed red) obtained directly from $X$. However, compared to the eigenvalues (dashed green) obtained from the covariance matrix $X^TX$, we can see that there is a disagreement as the required relative precision in $X$ approaches $\sim10^{-10}$, corresponding to a relative precision of $\sim10^{-20}$ in the $X^TX$ computation which is not possible with the \texttt{float64} datatypes.}
    \label{fig:svd_numerics}
\end{figure}

This can be seen in Figure \ref{fig:svd_numerics} where the PCA eigenvalues (rather their square roots) should, in principle, correspond with the data RMS seems to deviate starting at the precision limit. This is an indication that the underlying eigenvectors are numerically inaccurate. Compared with the eigenvalues from SVD, the data RMS agrees well with the singular values as expected. 

Performing singular value decomposition is much more stable numerically since it involves computation on the data $\X$ directly. However, since the products of SVD also involve the very large $\V$ matrix of shape $(\texttt{npix}\times\texttt{npix})$, the memory requirements for this step can be substantial. 

\section{Intrinsic variation of SINF}
\label{app:sinf_seed}
The total error in our measurements receives contributions from three different sources: the inevitable net bias in the method (it does not matter as long as it is smaller than the noise), the noise coming from the noise in the underlying data, and the stochastic noise associated with the random seed internal to SINF.
For each run, the Normalizing Flow code \texttt{SINF} converges slightly differently due to the intrinsic random nature of the algorithm mapping the base distribution to the target distribution. For a fixed noise realization, the spread in likelihood is smaller than, if not comparable to, the variation due to different noise realizations. We separate three sources of noise as follows. For the same data, we vary both the SINF and the random seed. The mean and the variance in output variation correspond to the total bias and the total noise. In a separate set of runs, the SINF seed is at the fixed noise realization seed and the variance of the output is calculated, which we associate with the SINF error. By subtracting this variance from the total variance, we can isolate the variance associated with the actual noise. Note that SINF variance is in general small and in any case a reducible variance.

\begin{figure}
    \centering
    \includegraphics[width=\linewidth]{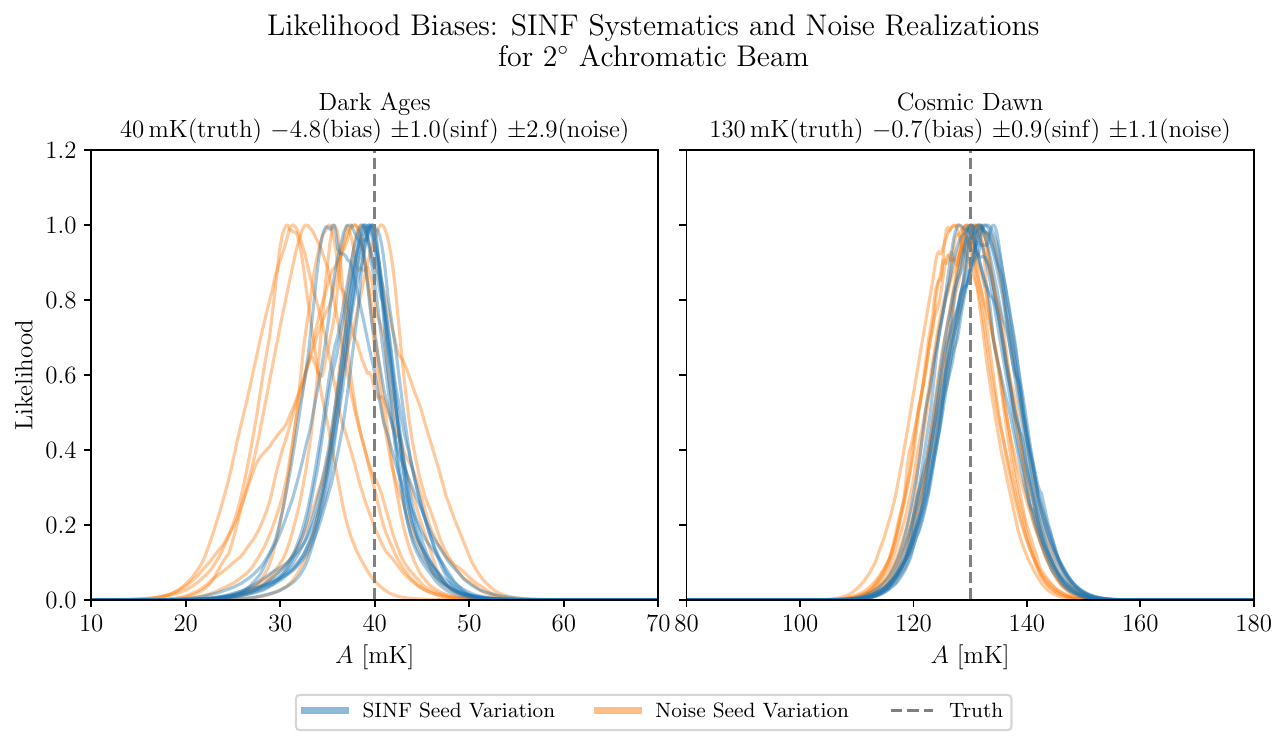}
    \caption{Variations due to different noise realizations (orange) and due to different random seeds in the SINF algorithm (blue) for an achromatic $2^\circ$ beam with $10^9$ SNR for the Dark Ages signal and $10^6$ SNR for the Cosmic Dawn signal. The spread in likelihood due to different noise realizations is larger than the spread due to different random seeds.
    \label{fig:intrinAchromatic}}
\end{figure}

\begin{figure}
    \centering
    \includegraphics[width=\linewidth]{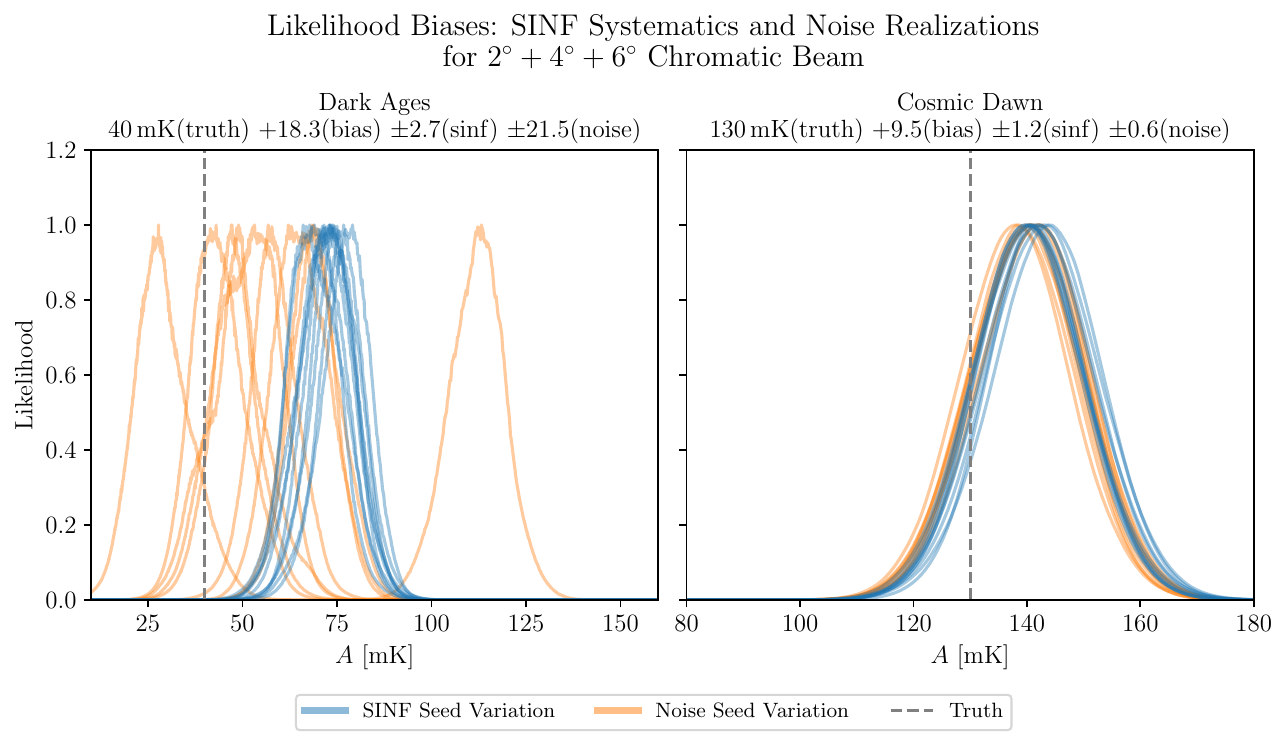}
    \caption{Variations due to different noise realizations (orange) and due to different random seeds in the SINF algorithm (blue) for a chromatic $2^\circ + 4^\circ + 6^\circ$ beam with $10^{10}$ SNR for the Dark Ages signal and $10^7$ SNR for the Cosmic Dawn signal. The spread in likelihood due to different noise realizations is larger than the spread due to different random seeds.
    \label{fig:intrinChromatic}}
\end{figure}

Shown in Figure \ref{fig:intrinAchromatic} for $\sigma=2^\circ$ achromatic beams and in Figure \ref{fig:intrinChromatic} for $2^\circ+4^\circ+6^\circ$ chromatic case for fixed SNRs. This exercise also guarantees that the SINF algorithm is sufficiently confident of its density estimation and converges stably.
\section{Effect of Galactic Cuts}
\label{app:galcut}

\begin{figure}
    \centering
    \includegraphics[width=\linewidth]{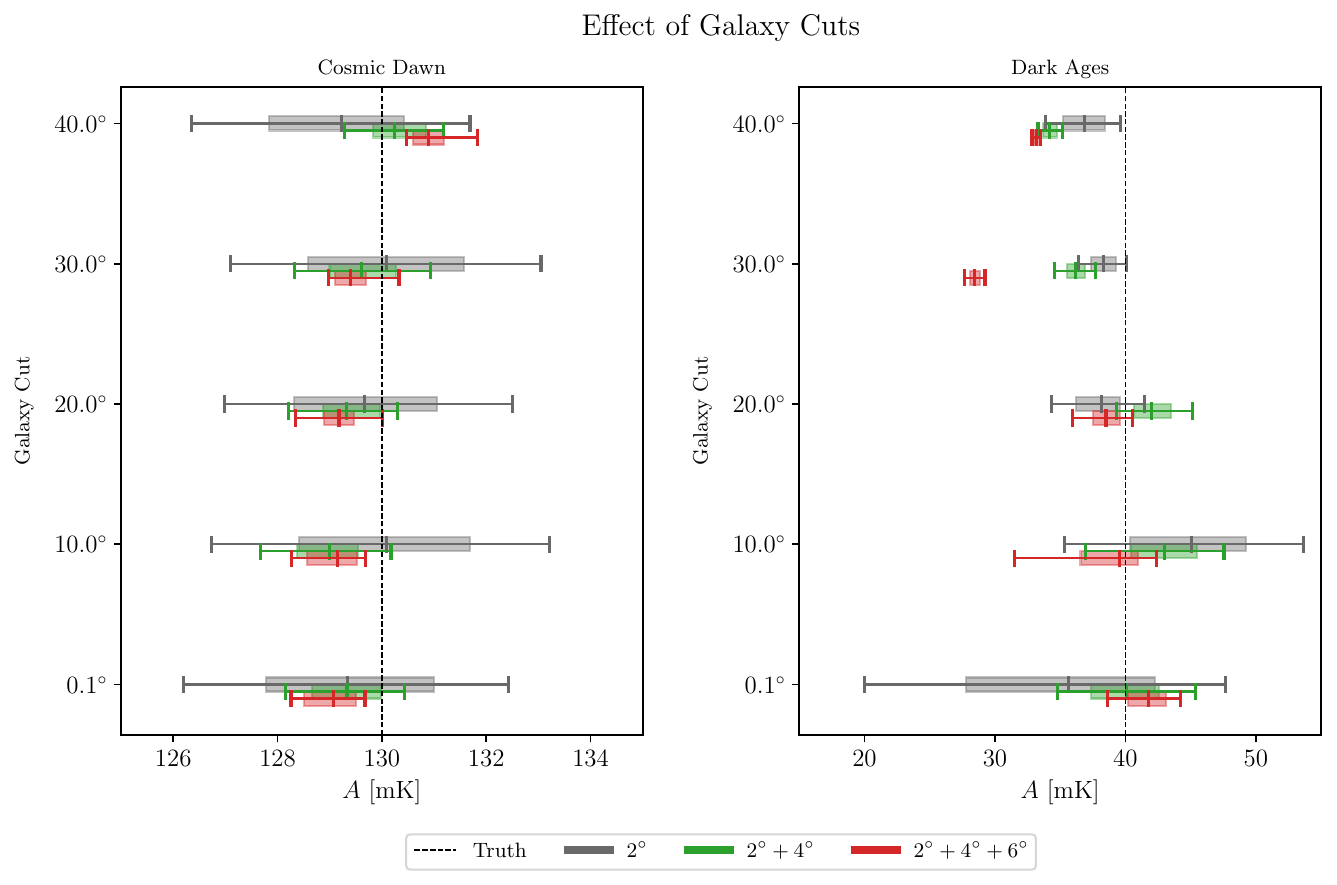}
    \caption{Effect of Galactic cuts on the likelihood constraints for the Dark Ages and Cosmic Dawn signals for achromatic beams. Box plot boxes represent 1-sigma errors, median line in the center, and whiskers bound the 95\% confidence intervals for each case. All results correspond to the idealistic, noiseless case.
    \label{fig:galcut}}
\end{figure}

We have tested the effect of removing the Galactic plane from the data. We find that the results are not significantly affected by the presence of the Galactic plane for Cosmic Dawn; however, for Dark Ages, a galactic cut of $\sim{20}^\circ$ is optimal. The results for the ideal, noiseless case are shown in Figure \ref{fig:galcut}.

While it is not realistic to mask the galaxy directly from the sky map, this exercise is useful in understanding the effect of foreground complexity on signal extraction. The results show that foreground complexity is the main limiting factor in signal extraction and that the method may be robust to the presence of the Galactic plane at the cost of a slight increase in the error bars.

\end{appendix}

\bibliography{21cm.bib}

\end{document}